\documentclass[twocolumn, aps, prb, showpacs, preprintnumbers, amsmath, amssymb, a4paper, superscriptaddress]{revtex4-2}
\usepackage[T1]{fontenc}

\usepackage{mathrsfs}
\usepackage{amssymb}
\usepackage{amsmath}
\usepackage{graphicx}
\usepackage{braket}
\usepackage{physics}
\usepackage{bm}
\usepackage{bbold}
\usepackage[utf8]{inputenc}
\usepackage{color}
\usepackage[colorlinks=true,citecolor=blue,urlcolor=blue]{hyperref}
\begin{document}
\preprint{APS/123-QED}

\title{One-dimensional topological channels in heterostrained bilayer graphene}
\author{Nina C. Georgoulea}
\affiliation{School of Physics, AMBER \& CRANN Institute, Trinity College Dublin, Dublin 2, Ireland}
\author{Nuala M. Caffrey}
\affiliation{School of Physics, University College Dublin, Dublin 4, Ireland}
\affiliation{Centre for Quantum Engineering, Science, and Technology, University College Dublin, Dublin 4, Ireland}
\author{Stephen R. Power}
\email{stephen.r.power@dcu.ie}
\affiliation{School of Physics, AMBER \& CRANN Institute, Trinity College Dublin, Dublin 2, Ireland}
\affiliation{School of Physical Sciences, Dublin City University, Glasnevin, Dublin 9, Ireland}

\date{\today}

\begin{abstract}

The domain walls between AB- and BA-stacked gapped bilayer graphene have garnered intense interest as they host topologically-protected, valley-polarised transport channels. The introduction of a twist angle between the bilayers and the associated formation of a Moir{\'e} pattern has been the dominant method used to study these topological channels, but heterostrain can also give rise to similar stacking domains and interfaces.
Here, we theoretically study the electronic structure of a uniaxially heterostrained bilayer graphene. We discuss the formation and evolution of interface-localized channels in the one-dimensional Moir\'{e} pattern that emerges due to the different stacking registries between the two layers. 
We find that a uniform heterostrain is not sufficient to create one-dimensional topological channels in biased bilayer graphene. 
Instead, using a simple model to account for the in-plane atomic reconstruction driven by the changing stacking registry, we show that the resulting expanded Bernal-stacked domains and sharper interfaces are required for robust topological interfaces to emerge. 
These states are highly localised in the AA- or SP-stacked interface regions and exhibit differences in their layer and sublattice distribution depending on the interface stacking. 
We conclude that heterostrain can be used as a mechanism to tune the presence and distribution of topological channels in gapped bilayer graphene systems, complementary to the field of twistronics.

\end{abstract}
\maketitle

\section{Introduction}
Graphene was the first 2D material to be experimentally isolated, and monolayer graphene (MLG) remains the most well-known 2D material due to a range of unique mechanical and electronic properties~\cite{novoselov2004electric, novoselov2005two, katsnelson2007graphene, lee2008measurement}. 
MLG has gapless, linear bands which form two Dirac cones, or valleys, near the Fermi energy~\cite{wallace1947band, castroneto:graphenereview}. 
The honeycomb lattice of MLG can be divided into two triangular sublattices (A and B), which give rise to the pseudospin degree of freedom underpinning exotic phenomena such as Klein tunneling~\cite{katsnelson2006chiral}. 
Due to its high elasticity and flexibility, it has also been suggested for potential applications in flexible electronic devices~\cite{novoselov2012roadmap}, including touch screens~\cite{bae2010roll} and foldable organic light emitting diodes~\cite{han2012extremely}.
It is also the building block for more complex multilayer structures~\cite{geim2013van}, such as bilayer graphene (BLG), which is formed from two stacked graphene layers coupled by weak van der Waals interactions~\cite{mccann2013electronic}. It can also be used in heterostructures, for example by combining graphene with transition metal dichalcogenides~\cite{azadmanjiri2020graphene}.

\begin{figure}[tphb]
    \center
    \includegraphics[width=.98\linewidth]{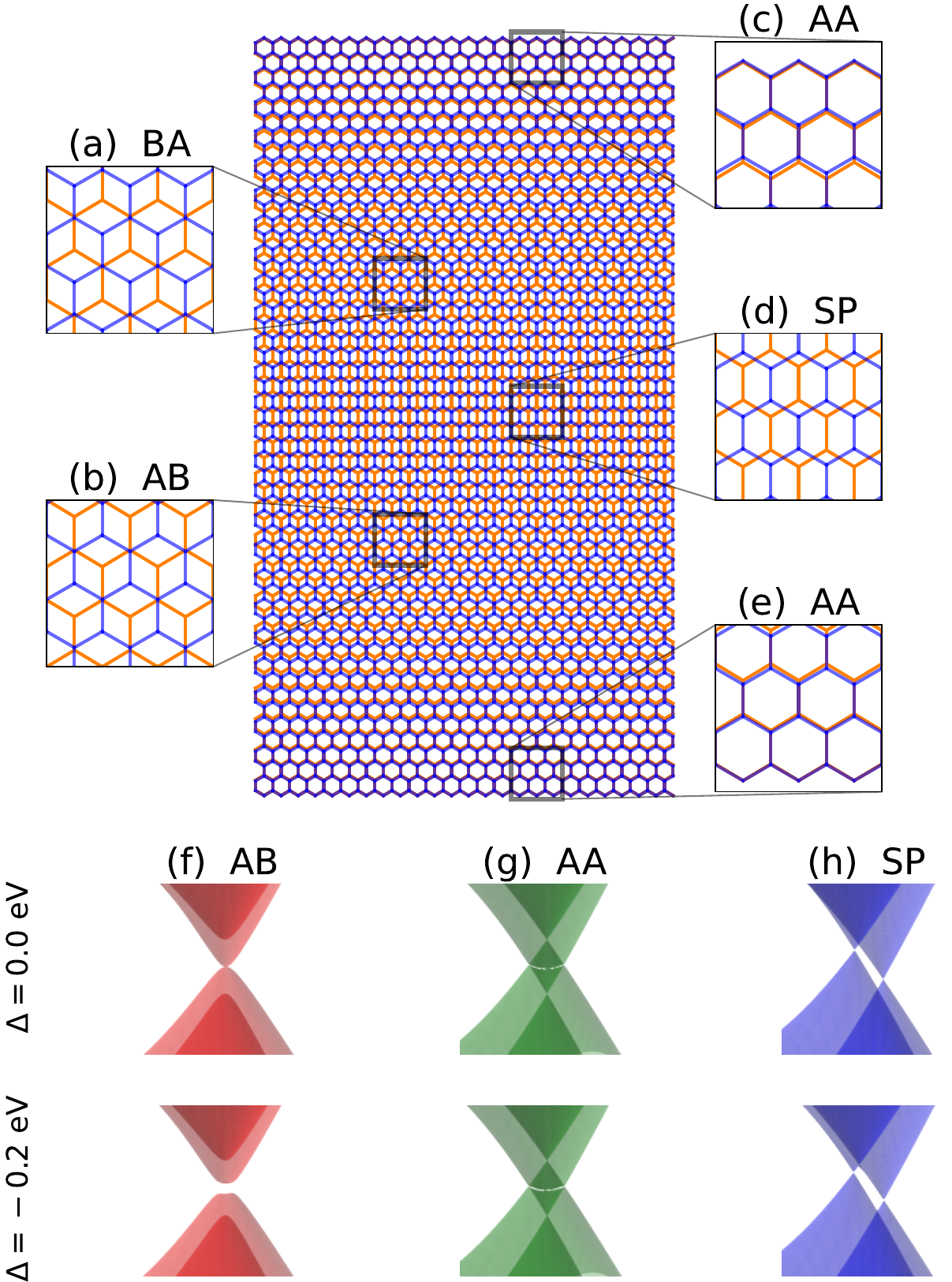}
    \caption{Bilayer graphene with a 4\% uniform heterostrain, applied in the AC-direction to the bottom (orange) layer (L1). A one-dimensional Moir\'{e} pattern emerges due to different stacking registries between the two layers, shown by the zoomed regions in panels (a)--(e). The band structures for BLG systems, with and without an interlayer bias, for uniform (f) AB-, (g) AA- and (h) SP-stacking registries are shown in the bottom panels.} 
    \label{fig:stackings}
\end{figure}

BLG systems can be characterised by the relative stacking between their layers.
In the most favourable stacking, the carbon atoms from one sublattice in each layer lie directly opposite the centre of a hexagon in the other layer, whereas the atoms from the remaining sublattice in each layer lie directly opposite each other ~\cite{mccann2013electronic, rozhkov2016electronic}, forming so-called `dimers'.
There are two equivalent possibilities, AB and BA stacking, depending on which sublattices from each layer form dimers, as shown in Fig.~\ref{fig:stackings}(a) and (b).  
The least favourable stacking possibility is AA-stacking, where the two layers are perfectly aligned and each atom sits opposite an atom from the same sublattice in the opposite layer, as in Fig.~\ref{fig:stackings}(c) and (e).
Between the low-energy AB and high-energy AA extremes lie a range of stacking options which can be achieved by varying the amount and directions by which one layer is shifted relative to the other.
The sublattice symmetric SP (or saddle point) stacking, shown in Fig.~\ref{fig:stackings}(d) is of interest in this work as it occurs for shifts half way between AB and BA stackings, and corresponds to a local energy maximum along this direction.

The electronic properties of a BLG structure depend sensitively on the stacking sequence, as shown in Fig.~\ref{fig:stackings}(f)--(h). 
For example, AB or BA-stacked BLGs have parabolic electronic bands, whereas the bands of AA-stacked BLGs remain linear, as in MLG~\cite{ho2006coulomb, novoselov2004electric}, but the two resultant cones are split in energy.
SP-stacked BLGs have also linear bands, but the resultant cones now have relative shifts in both energy and momentum~\cite{park2015electronic}.
These differently stacked systems also behave very differently under the application of an interlayer bias.
A band gap proportional to the bias is opened for AB/BA stackings~\cite{castro2007biased}, while AA- and SP-stacked systems remain metallic~\cite{silva2020electronic}. 

The gaps opened by an interlayer bias in AB- and BA-stacked BLGs are equal in magnitude, but are topologically non-equivalent ~\cite{zhang2013valley, san2013helical}.
Interfaces between AB and BA-stacked regions in a single system are therefore of interest as they can host topologically-protected, valley-polarised transport channels.
These appear in the band structure of biased BLG systems as pairs of bands (in each valley) which connect the otherwise gapped Dirac cones~\cite{zhang2013valley, vaezi2013topological, ju2015topological}. 
Interfaces between AB- and BA-stacked regions, and the associated topological channels, can occur in a number of systems, with the simplest example being a change in stacking registry caused by a grain boundary in one of the layers~\cite{yasaei2014chemical, jaskolski2016existence}.
A similar effect occurs if the sign of the bias, instead of the stacking registry, changes across the interface~\cite{oostinga2008gate, zhang2009direct, li2016gate, rickhaus2018transport}. 
AB/BA interfaces also occur when a relative twist angle is introduced between the two layers, to form twisted bilayer graphene (tBLG). 
This creates a Moir{\'e} pattern where the stacking order has a periodic modulation, with alternating domains of AA, AB/BA and SP stackings~\cite{bistritzer2011moire, andrei2020graphene}.
The interfaces between AB and BA domains form a triangular superlattice with AA-stacked vertices connected by SP-stacked edges. 
The formation of localised, topological interface states along the SP edges when a bias is applied is a topic of huge interest recently, both theoretically~\cite{san2013helical, efimkin2018helical, fleischmann2019perfect, tsim2020perfect, de2021network} and experimentally~\cite{PhysRevLett.121.037702, kerelsky2019maximized, xu2019giant, verbakel2021valley}.
The localisation of electronic states in different domains can lead to correlated insulating states and unconventional superconductivity for specific twist angles~\cite{cao2018unconventional, cao2018correlated, stepanov2020untying}.

Strain can also break the stacking registry and give rise to different stacking domains and interfaces between them.
A heterostrain, i.e. a difference between the strains felt by each layer, introduces a mismatch between the lattice constants in each layer.
Local compression, stretching or slipping of one layer can lead to the formation of strain solitons or more complex topological defects~\cite{popov2011commensurate, alden2013strain}.
A pure uniaxial heterostrain, arising from a uniaxial strain present in only one of the layers, leads to a one-dimensional (1D) Moir{\'e} pattern~\cite{choi2010controlling, van2016piezoelectricity, georgoulea2022strain, schleder2023onedimensional}.
An example is shown in the main panel of  Fig.~\ref{fig:stackings}, where a 4\% armchair-direction uniaxial tensile strain  is present in the bottom (orange) layer, with the top (blue) layer remaining unstrained.
The resulting lattice mismatch in this direction gives rise to an alternating, repeating sequence of (from bottom to top in Fig.~\ref{fig:stackings}) AA, AB, SP, and BA stackings.
Such a situation can arise, for example, if the strain applied to a bilayer system using a flexible substrate does not transfer between the layer directly in contact with the substrate to the top layer~\cite{androulidakis2020tunable, frank2012phonon, wang2019robust}.
In a recent work~\cite{georgoulea2022strain}, we demonstrated that such a situation can arise if the energetic penalty caused by the broken AB-stacking is less than the energy cost required to strain the top layer by the same amount.
For applied heterostrains of approximately 1\% or greater, we found that modulated stacking case is energetically favourable, which suggests that such strains can be used as a tool to tune the stacking of a BLG system.

In this work, we focus on the formation and evolution of interface states in heterostrained BLGs. 
The 1D symmetry of the system gives rise to two types of interface between AB and BA regions, namely those centred around the SP-stacked and the AA-stacked regions, as shown in Fig.~\ref{fig:stackings}.
In Sec.~\ref{sec:methods}, we introduce the tight-binding methods use to calculate the electronic properties of general BLG systems and a simple model to account for in-plane relaxations and changes in domain size and interface sharpness.
In Sec.~\ref{sec:3}, we examine the band structure and distribution of states in BLGs with a uniform heterostrain, and compare the results to what is seen for uniformly-stacked AA, AB and SP systems.
Realistic values of bias and strain do not give clear signatures of the expected topological interface channels, which we attribute to the relatively wide interface regions and the relatively small AB and BA domains that occur for a uniform heterostrain.
Following this, in Sec. \ref{sec:smoothing}, we use the simple model to mimic the effects of in-plane relaxation and the resulting variances in interface width.
This is motivated by previous studies in twisted systems, which find that relaxation tends to minimise AA-stacked domains, maximise AB/BA-stacked domains and sharpen SP-stacked interfaces~\cite{ popov2011commensurate, lin2013ac, van2015relaxation, dai2016twisted, jain2016structure, nam2017lattice, gargiulo2017structural, carr2018relaxation, lin2018shear, guinea2019continuum, yoo2019atomic, fleischmann2019perfect, lucignano2019crucial, tsim2020perfect, nguyen2021electronic, leconte2022relaxation}.
These structural deformations can affect the electronic band structure and, in particular, the formation of flat bands at specific twist angles.
In our heterostrained systems, one layer is assumed to remain uniform due to its interaction with a substrate. 
If strain is applied using a flexible substrate, the in-plane relaxation should occur in the unstrained layer~\cite{yu2008raman, huang2009phonon, anagnostopoulos2015stress, androulidakis2020tunable}.
However, if strain is applied to the top layer using, for example, the tip of an atomic force microscope~\cite{kapfer2022programming}, then we expect in-plane relaxation within this layer.
We therefore consider in-plane relaxations of both the strained and unstrained layers independently in order to capture the geometries that emerge from both processes.
We find that in-plane relaxation, of either the strained or unstrained layer, allows for more robust topological interfaces to emerge due to larger, gapped AB/BA domains and sharper SP and AA interfaces.
After discussing the structural, energetic and band-structure considerations, we examine the resulting topological interfaces in more detail in Sec. \ref{sec:interfaces}.
Finally, we discuss our findings in the context of recent experimental works and highlight the potential role of strain as a mechanism to tune the presence and distribution of topological channels in BLG systems.

\section{Methods}
\label{sec:methods}
\subsection{Electronic structure}
To model the electronic structure of a general multilayer graphene system with a range of different stackings, strains, inter-layer biases and/or twists, a tight-binding (TB) Hamiltonian of the form
\begin{equation}
\hat H = \sum_i \epsilon_i  \hat c^{\dagger}_{i}\hat c_{i} -t(\vec{d})\sum_{\langle i, j \rangle}\big(\hat c^{\dagger}_{i}\hat c_{j} + \hat c^{\dagger}_{j}\hat c_{i}\big)
\end{equation}
can be used.
Here $\hat c^{\dagger}_{i}$ and $\hat c_{}$ are, respectively, the creation and annihilation operators for an electron in the $p_z$ orbital at site $i$.
The effect of a simple interlayer bias $\Delta$ can be included by setting the onsite parameters $\epsilon_i = \pm\frac{\Delta}{2}$, with the choice of sign determined by the layer in which site $i$ is located.
The hopping parameter $t(\vec{d})$  depends on the distance vector $\vec{d} = \vec{R}_{i}-\vec{R}_{j}$ between the two sites and is written in terms of the appropriate Slater-Koster parameters~\cite{slater1954simplified}) as
\begin{equation}\label{eqq}
    -t(\vec{d}) = V_{pp\pi}(d)\Big[1-\Big(\dfrac{\vec{d}\cdot \vec{e}_{z}}{d}\Big)^{2}\Big] + V_{pp\sigma}(d)\Big(\dfrac{\vec{d}\cdot \vec{e}_{z}}{d}\Big)^{2}
\end{equation}
\begin{align}
V_{pp\pi}(d) & =V^{0}_{pp\pi}e^{-\tfrac{d-a_{0}}{r_{0}}} \\
V_{pp\sigma}(d) &=V^{0}_{pp\sigma}e^{-\tfrac{d-d_{0}}{r_{0}}}.
\end{align}
Here $V^{0}_{pp\pi} = -2.7$ eV and $V^{0}_{pp\sigma} = 0.48$ eV are, respectively, the nearest-neighbour in-plane and out-of-plane (dimer) transfer integrals for unstrained graphene systems, $a_0 = 1.42$ \AA\ and $d_0 = 3.35$ \AA\ are the corresponding unstrained in-plane and out-of-plane atomic separations, $r_0 = 0.453$ \AA\ is a decay length, and $\vec{e}_{z}$ is the unit vector in the out-of-plane $z$ direction.
We include all hopping terms whose corresponding separation in the $xy$-plane is less than a cutoff of $r_c = 1.82$ \AA . 
Within a single layer, this is equivalent to a nearest-neighbour TB model that can account for local strains up to $\epsilon  \sim 0.25$.
This value is also sufficient to account for the dimer ($\gamma_1$) and skew ($\gamma_3, \gamma_4$) interlayer hoppings, up to similar levels of deformation, in BLG systems.
This type of TB model is suitable for systems with inhomogeneous strains and stackings, as it can account for differing bond lengths in both the in-plane and out-of-plane directions, and has previously been applied in studies of twisted bilayers~\cite{trambly2010localization, nam2017lattice, tsim2020perfect, leconte2022relaxation}. 

The main geometry considered in this work is a BLG sheet with a uniaxial heterostrain of 1\% applied along the armchair direction in one of the layers (L1 -- the `bottom' layer), with the other layer (L2 -- the `top' layer) unstrained. 
As discussed in the introduction, the layer which is actually strained in experiment will depend on the method used to apply the strain.
In the absence of a Poisson contraction perpendicular to the applied strain, this creates a regular one-dimensional Moir\'e pattern.
This is shown for an exaggerated strain of 4\% in Fig. \ref{fig:stackings}.
As the stacking modulation does not affect periodicity in the zigzag direction, the periodic unit cell required for calculations of this system consists of a single chain of atoms from each layer.
We assume an initial AA-stacking, which remains unchanged at $y=0$ and therefore the stacking must return to AA at the top of the unit cell in order to maintain periodicity in the $y$-direction.
This requires a total extension which is a multiple of $y_u = 3 a_{0}$, the width of a four-atom graphene unit cell.
If L1 contains $N$ strained cells and has a total extension of $M y_u$, then L2 contains $N+M$ unstrained cells.
This restricts calculations to rational strain values $\varepsilon = \frac{M}{N}$ and requires larger unit cells to simulate smaller strains. 
The unit cell for the 1\% heterostrain case discussed throughout this work corresponds to $M=1$, $N=100$ and contains 804 carbon atoms.
The band structures presented in the following sections are calculated using exact diagonalization of the corresponding Hamiltonian matrices with the appropriate Bloch phases.
As we are largely concerned with the emergence of states confined to $x$-direction interfaces, we only show bands as a function of $k_x$, with $k_y=0$.

\subsection{Simple model of relaxation}
\label{sec:simple}
\begin{figure}[]
    \center
    \includegraphics[width=.98\linewidth]{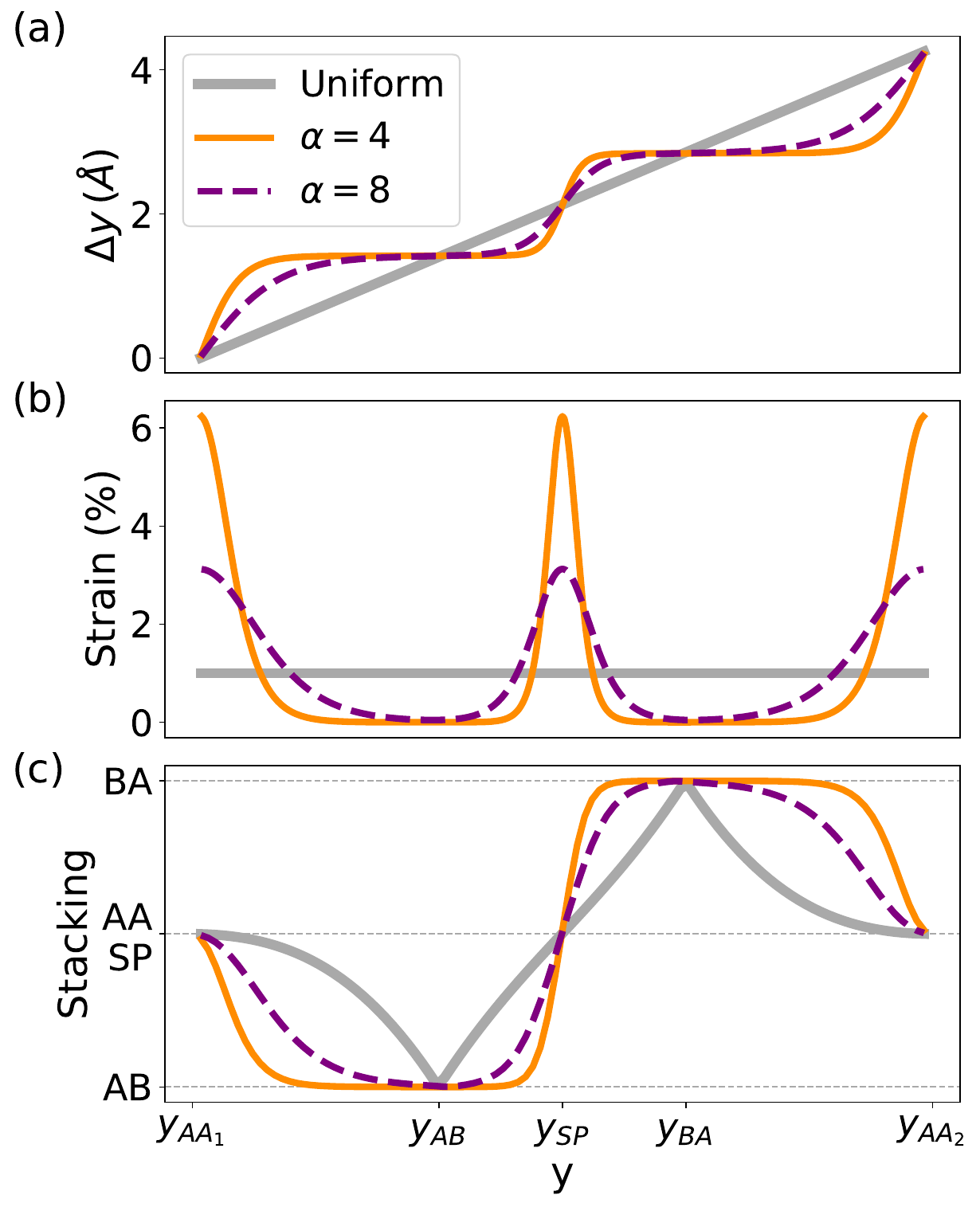}
    \caption{Comparison of the local (a) displacement, (b) strain and (c) stacking order in heterostrained bilayer graphene as a function of position along the strained direction ($y$). The grey curve shows the case of a uniform uniaxial strain in layer L1, whereas the orange and purple curves correspond to `relaxed' structures with different values of the interface smoothness parameter $\alpha$.} 
    \label{fig:relax}
\end{figure}

In-plane relaxation has been shown to play a vital role in determining the electronic properties of twisted Moir\'e systems~\cite{nam2017lattice, lin2018shear, fleischmann2019perfect, guinea2019continuum, tsim2020perfect, nguyen2021electronic}, with the main effect of such relaxation an increase in the size of AB and BA domains at the expense of AA and SP regions.
In the presence of an interlayer bias, this affects the localization of states along the sharper SP interfaces between gapped regions of opposite mass. 
To consider how in-plane relaxations could affect our heterostrained geometry, we employ a simple model that allows the sharpness of the AA- and SP- interfaces, and hence the effective width of AB/BA regions, to be continuously tuned.
In our model, we assume that only one layer of the system is allowed to relax. 
We first consider the case where relaxation occurs within L1, the strained layer. 
A uniform $y$-direction strain $\varepsilon = \frac{1}{N}$ applied to L1 produces an extension which is linearly-dependent on the unstrained y-coordinate $y_{0, L1}$:
\begin{equation}
\Delta y_{L1} \, (y_{0, L1}) = \varepsilon \, y_{0, L1} \,.
\label{eq:simple_strain}
\end{equation}
The linear extension and uniform strain, as a function of position, are shown by the grey curves in Fig.~\ref{fig:relax}(a) and (b) respectively.
The resultant lattice mismatch with L2 (the unstrained layer) gives rise to high-symmetry AA, AB, SP and BA stackings when $\Delta y_{L1} = 0$, $a_0$, $\tfrac{3}{2} a_0$ and $2 a_0$ respectively, which correspond to $y_{0, L1}^{AA_1} = 0$, $y_{0, L1}^{AB} = \frac{N y_u}{3}$, $y_{0, L1}^{SP} = \frac{N y_u}{2}$ and $y_{0, L1}^{BA} = \frac{2 N y_u}{3}$.
The choice of a rational strain values, as discussed above, ensures that the stacking returns to AA again at $\Delta y_{L1} = y_u = 3 a_0$ and $y_{0, L1}^{AA_2} = N y_u$.
(Note that, due to periodicity, AA$_1$ and AA$_2$ correspond to the same point, but it is useful to consider them separately in the context of a finite unit cell.)
The positions of these high-symmetry stacking locations are shown by the grey curve in Fig.~\ref{fig:relax}(c) and the local stacking near them by the insets (a)--(e) in Fig.~\ref{fig:stackings}.

To simulate sharper interfaces, and broader gapped regions, we replace the linear extension from Eq. \eqref{eq:simple_strain} with a sum of sigmoid functions, $\Delta y_{L1}^{\mathrm{rel.}} \, (y_{0, L1}) = \sum_i S_i - \tfrac{S_{{AA}_1}}{2}$, each of the form
\begin{equation}
    S_i(y_{0, L1})=\dfrac{\Delta Y_i}{1+e^{-(y_{0, L1}-y_{0, L1}^i)/(\alpha \Delta Y_i )} }  \,,
    \label{eq:sigmoid}
\end{equation}
with $i \in \{ AA_1, SP, AA_2 \}$.
These `S'-shaped functions are centred at the AA and SP interfaces (see orange and purple curves in Fig.~\ref{fig:relax}(a)), where they locally increase the strain (panel (b)) and sharpen the interfaces (panel (c)). 
This leads to much lower strains, and more uniform stackings, in the AB and BA regions.
The `smoothness' or `sharpness' of an interface is determined by the parameter $\alpha$, which also sets the local maximum value of strain at the interface. 
$\alpha$ relates the characteristic width of an interface to the total stacking shift ($\Delta Y_i$) that occurs across it. 
We note that the AA-interface has twice the interlayer shift of the SP-interface ($\Delta Y_{AA} = 2a_0$, $\Delta Y_{SP}=a_0$), but that $\alpha$ acts as a single adjustable parameter to allow relaxation, with similar levels of strain, in both interfaces.
Low values of $\alpha$ give sharp interfaces, with more uniform strain returning as $\alpha$ is increased.
However, the model assumes that the effects of each interface are independent and breaks down if $\alpha$ is increased to values where the interface regions begin to merge.

The above approach can also be used to consider relaxation occurring instead in the initially unstrained layer (L2), as may be expected to occur if strain is applied to the other layer via a flexible substrate.
In this case, a uniform tensile strain is first applied to L2 to match that in the strained layer, before a non-uniform compressive strain is applied using the `sum-of-sigmoids' approximation above to return the average strain in the unstrained layer to zero.  
The structural and electronic effects of both types of relaxation are discussed in detail in Sec. \ref{sec:smoothing}.

\section{Electronic properties of heterostrained systems}
\label{sec:3}
\begin{figure}[tpb]
    \center
    \includegraphics[width=.95\linewidth]{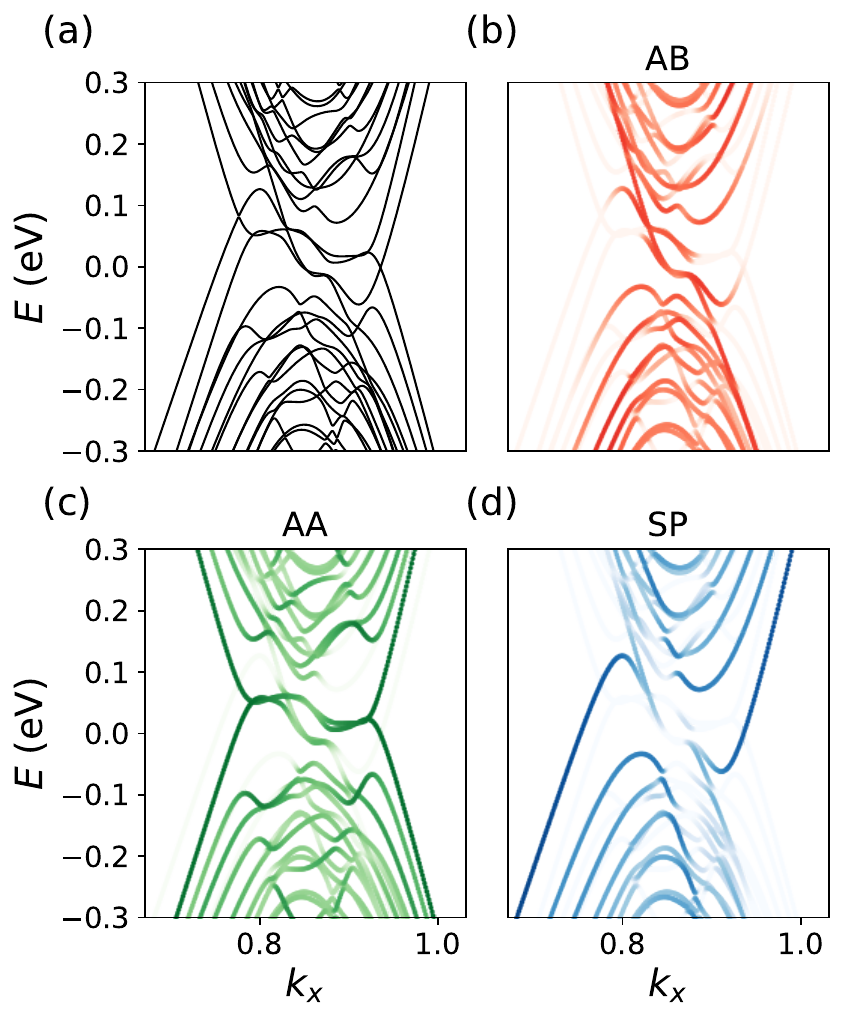}
    \caption{(a) $K$-valley band structure of a BLG system with 1\% uniform heterostrain. Panels (b), (c), (d) show the band structure projected onto 2.0 nm wide strips around the perfect AB-, AA- and SP-stacking points, respectively. Most of the states are distributed across multiple regions, with exceptions for low-energy, high-$k$ states which show confinement to regions with particular stackings.} 
    \label{fig:bands}
\end{figure}

We begin by considering the electronic structure of BLG with a 1\% strain applied to one of the layers, in the absence of either interlayer potentials or relaxation effects.
The 804-atom ($43$ nm wide) unit cell of this system is four times wider than the schematic structure shown in Fig. \ref{fig:stackings}, with a minimum separation of approximately 14 nm between the centres of the AB and BA domains.
The electronic structure of this system along the $k_x$ direction is shown in Fig. \ref{fig:bands}(a) and reveals a complex series of subbands due to the large real-space unit cell and associated band-folding in reciprocal space.
The parabolic bands and semi-metallic, zero-bandgap behaviour expected for AB-stacked BLG, shown schematically in Fig. \ref{fig:stackings}(f), cannot be clearly distinguished here. 
Instead, a large number of bands cross the Fermi energy near $E=0$ and give the system a metallic character.
This suggests that modulated stacking introduces features from the band structures of uniformly AA (Fig. \ref{fig:stackings} (g)) and SP-stacked (Fig. \ref{fig:stackings} (h)) BLGs.
Unlike AB-stacking, these cases resemble the conical band structure of MLG, but with two copies of the linear MLG cone separated in either energy (AA), or energy and momentum (SP). 

We now explore the relationship between the band structure of the heterostrained structure and those of regularly-stacked systems in more detail.
Fig. \ref{fig:bands}(b--d) show the bands from Fig. \ref{fig:bands}(a) projected onto narrow 2.0 nm strips centred around $y^{AB}$, $y^{AA}$, and $y^{SP}$, with the colour of each point showing the weight of the associated heterostrained bilayer state in these regions. 
These plots reveal the effective band structures in the AB, AA and SP regions of the heterostrained BLG.
Bands which are only strongly coloured in one region indicate states that are largely localised in that region.
Similarly, a strong correspondence between the bands in each region and their bulk counterparts in Fig.~\ref{fig:stackings}(f)--(h) would indicate that bands in a region with a particular stacking resemble those in a uniform bilayer with the same stacking.
The majority of the heterostrained systems states, and particularly the higher-order subbands, are not localised in this manner.
Instead they have weight in multiple regions, indicating states that are distributed across large parts of the system and not confined to regions with a particular stacking.
These states reside in regions of $k_x$--$E$ common to all three stackings, and can be viewed as hybridisations between states with similar momentum and energy that occur in regions with different stackings. 
The only significant exceptions to this trend is for states outside the region of $k_x$--$E$ space spanned by the bands of AB-stacked BLG, e.g., larger $k$ values at low energies.
These states are highly quenched in AB regions, preventing the hybridization of states in different regions, and thus leading to bands that are highly-localised in either AA or SP regions. 
Although they have a similar spatial distribution to that expected for topological interface states, we note that there is no gap in this system, and that these states emerge instead from simple confinement effects.

\begin{figure}[tpb]
    \center
    \includegraphics[width=.95\linewidth]{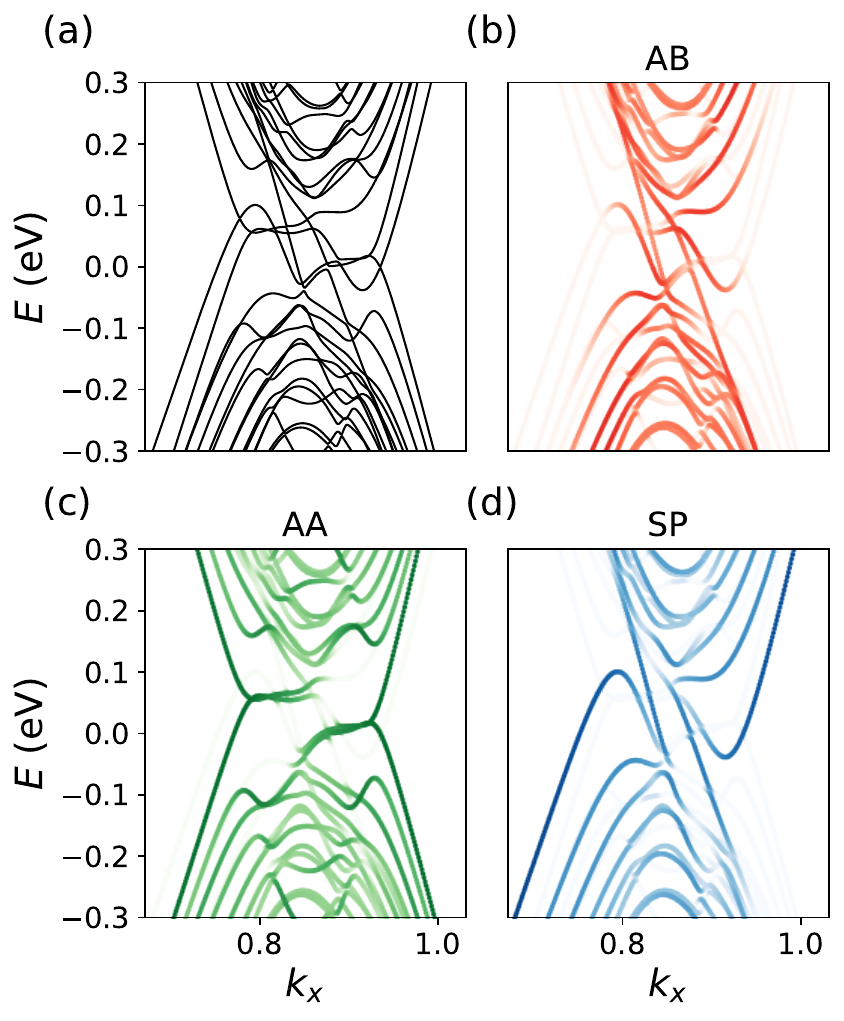}
    \caption{Similar to Fig. \ref{fig:bands}, but with an additional interlayer bias of $\Delta=-200$ meV between the layers. No clear band gap opens, even for states in the AB-stacked region, despite a sizeable band gap opening for the corresponding uniformly AB-stacked system in Fig. \ref{fig:stackings}(f). }
    \label{fig:bands_bias}
\end{figure}

An interlayer bias opens a band gap only in BLG systems with certain stackings, with the largest gap occurring for AB-stacking and no gap opened for AA- or SP-stacked systems. 
This leads to a distribution of gapped and conducting domains, and the formation of topological interface states, in systems with modulated stacking, such as tBLG~\cite{san2013helical}.
Similar behaviour should emerge in heterostrained structures and, in particular, the AA and SP strips separating AB and BA regions should host one-dimensional topologically protected channels. 
Fig. \ref{fig:bands_bias} shows the band structure and projections of the system from Fig. \ref{fig:bands} when an interlayer bias of $-200$ meV is applied. 
We note that here the negative potential is applied to the strained layer -- a slightly different result is obtained for the opposite case due to the strain-induced layer asymmetry and the broken electron-hole symmetry resulting from longer-ranged hopping terms in the Hamiltonian. 
This bias is sufficient to open a clear gap for a uniform AB-stacked structure (Fig. \ref{fig:stackings}(f)), but has little qualitative effect on the gapless bands of AA- and SP-stacked systems (Fig. \ref{fig:stackings}(g,h)).
We note that, unlike other one-dimensional interface cases such as grain boundaries or sharp bias flips~\cite{yasaei2014chemical, jaskolski2016existence, oostinga2008gate, zhang2009direct, li2016gate, rickhaus2018transport}, the application of an interlayer bias in Fig. \ref{fig:bands_bias} does not give rise to clearly defined topological channels.
In these other systems, an interlayer potential opens a bulk band gap which is bridged only by pairs of chiral boundary modes with opposite propagation directions in the K and K$^\prime$ valleys ~\cite{zhang2013valley}.
For our system, this would correspond to a total of four valley-protected topological modes in each valley, two each for the interfaces along the AA and SP domains. 
Instead of the clear emergence of a bulk gap and topological modes, we instead see only minor changes compared to the unbiased system discussed in Fig.~\ref{fig:bands}.

\section{Role of interface smoothness}

\label{sec:smoothing}
\begin{figure}[tpb]
    \center
    \includegraphics[width=.95\linewidth]{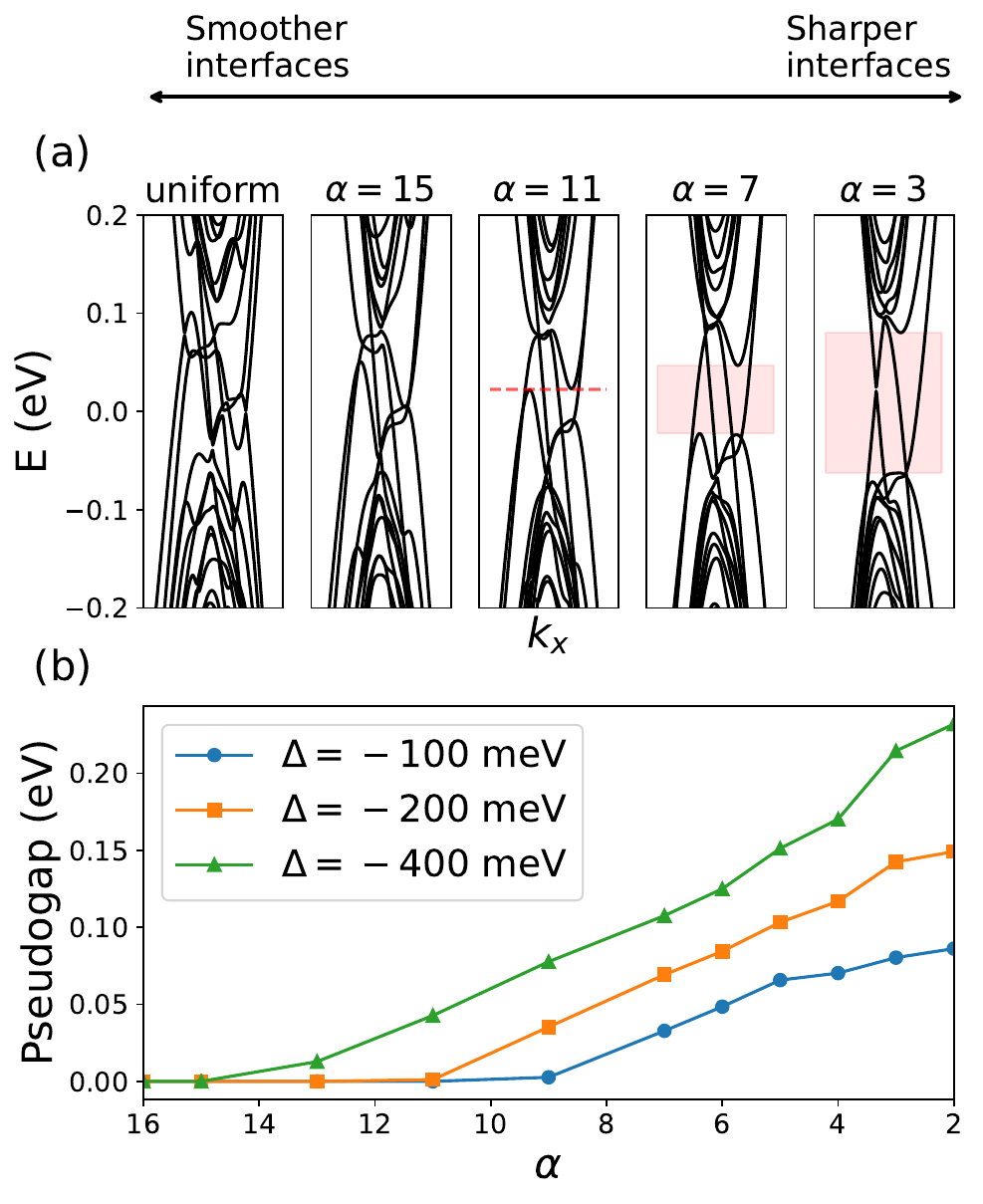}
    \caption{(a) Evolution of the band structure of the $\Delta=-200$ meV heterostrained BLG system as the AB and SP interfaces are sharpened by inhomogeneously relaxing the strained layer. The `pseudogap' window, where only four band crossings are present, is show in red. 
    (b) Pseudogap magnitude for different values of interlayer bias ($\Delta$) and interface smoothness ($\alpha$).}
    \label{fig:alphas}
\end{figure}

The absence of gap formation and topological modes in the previous section can be explained by the continuously varying stacking order in the system considered.
This results in only very small regions with perfect AB or BA stackings and prevents the formation of a band gap.
However, the continuously varying stacking order is an artefact of the homogeneous strain applied to L1, and is not necessarily representative of realistic systems where non-uniform strain profiles are adopted to minimise the total energy of the system~\cite{popov2011commensurate, lin2013ac, van2015relaxation, dai2016twisted, jain2016structure, nam2017lattice, gargiulo2017structural, carr2018relaxation, lin2018shear, guinea2019continuum, yoo2019atomic, fleischmann2019perfect, lucignano2019crucial, tsim2020perfect, nguyen2021electronic, leconte2022relaxation}.
We now consider the simple relaxation model from Sec. \ref{sec:simple} to determine how in-plane relaxation could affect the electronic properties of a biased, heterostrained BLG system.
This allows us to continuously adjust the interface smoothness and redistribute the strain from the AB/BA-stacked regions to the SP and AA interfaces. 
Fig. ~\ref{fig:alphas}(a) shows the evolution of the band structure of the biased system from the previous section as the interface smoothness $\alpha$ is decreased. 
As the interfaces become sharper, the number of bands crossing the Fermi energy decreases until only two pairs of crossings remain.
This is the expected number of band crossings for a BLG system with two topological interfaces between AB and BA-type regions, i.e., with the formation of counter-propagating valley-polarised interfaces states along both the AA and SP strips in our system~\cite{zhang2013valley}.
The energy window in which only these four bands are present, which we refer to as the \emph{pseudogap}, is shown by the red shaded areas for $\alpha = 7$ and $\alpha = 3$ in Fig.~\ref{fig:alphas}(a), and increases as the interfaces become sharper. 
The pseudogap only opens below a critical value of $\alpha = 11$, as shown by the red dashed line, and is not present for unrelaxed or smoother interfaces.
This evolution of the pseudogap is shown explictly by the orange curve in Fig.~\ref{fig:alphas}(b), together with corresponding results for smaller and larger values of the interlayer bias.
The critical smoothness required to open a pseudogap and observe interface states varies with the magnitude of the bias, with smaller biases requiring a sharper interface to open a pseudogap.
This is consistent with the different stacking profiles in these systems.
Sharper interfaces give rise to larger regions with nearly perfect AB or BA stacking where a smaller bias is able to effectively quench bulk states. 
Although precise AA or SP stackings still occur at the sharper interfaces, they do not persist over wide enough areas for the gapless bulk-like bands associated with uniform AA- or SP-stacked systems to emerge.

\begin{figure}[tpb]
    \center
    \includegraphics[width=.95\linewidth]{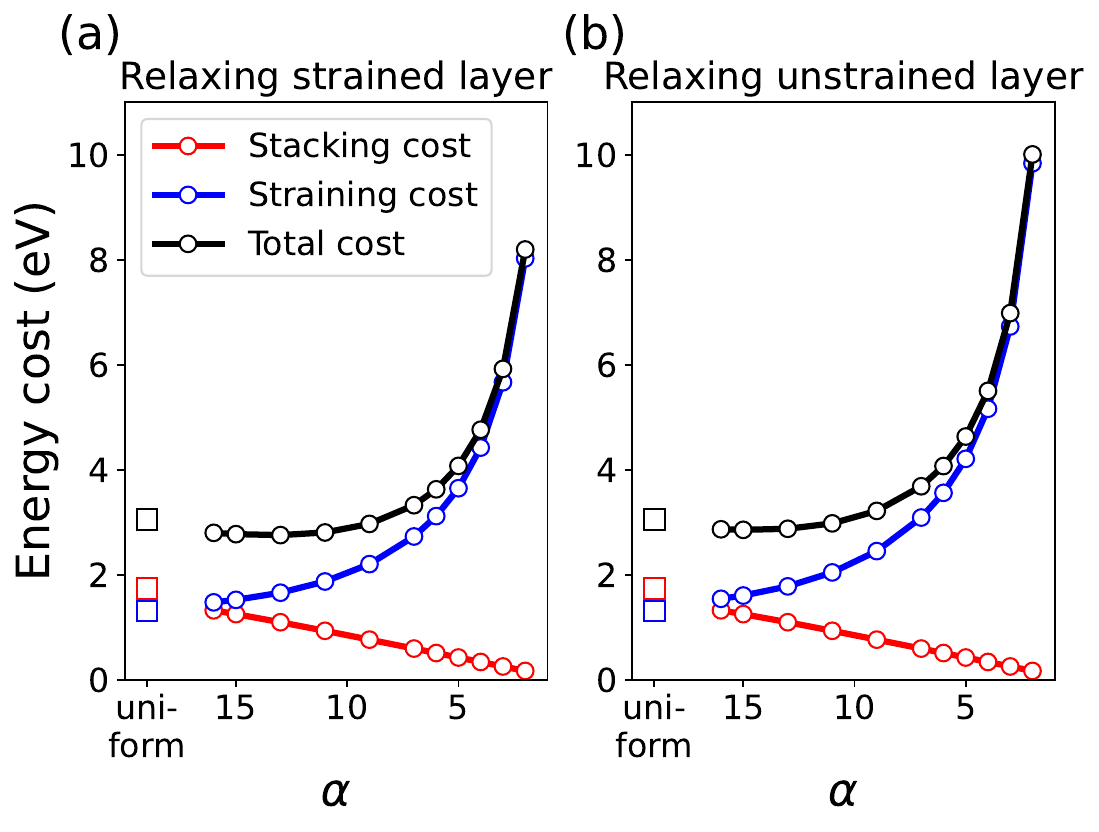}
    \caption{Energy costs (per supercell) associated with breaking AB-stacking registry (red curves) and strain (blue curves) in heterostrained systems. The panels show the effect of the simple relaxation model applied to either the (a) strained or (b) unstrained layer to give sharper interfaces. The black curves show the total energy from both terms, and the square symbols show the corresponding values for uniformly strained (unrelaxed) systems.} 
    \label{fig:energetics}
\end{figure}

The degree to which intralayer relaxation will occur can be simulated in a number of ways, including molecular dynamics~\cite{lin2013ac, fleischmann2019perfect, jain2016structure, leconte2022relaxation} or ab-initio simulations~\cite{lucignano2019crucial}, and the Frenkel Kontorova model and related continuum approaches~\cite{popov2011commensurate, nam2017lattice, carr2018relaxation, tsim2020perfect}.
The specifics of the relaxation are also likely to be affected by experimental conditions, such as substrate effects and the manner in which strain is applied.
While the specifics may vary, relaxation occurs principally to minimise the energy costs introduced by straining the graphene lattice and breaking uniform AB-stacking registry.
Both contributions can be calculated on a per-atom basis from ab-initio simulations of uniform systems~\cite{georgoulea2022strain}, allowing the energetics of larger, non-uniform systems to be estimated by summing over local stacking and strain costs throughout the system.
Inspired by the possible experimental processes, we consider two possible relaxation scenarios in Fig. \ref{fig:energetics}: either (a) the heterostrained layer or (b) the unstrained layer relaxes under the simple model in Sec. \ref{sec:simple}.
In each case, the atomic positions in the other layer are unaffected by the relaxation. 

Fig. \ref{fig:energetics} shows the energy costs (per supercell) due to breaking uniform AB-stacking (red curves) and due to strain (blue curves) in heterostrained systems with the relaxation model applied.
The total energy cost is shown by the black curve, with the corresponding energy values for a uniformly heterostrained system (without relaxation) shown by the square symbols. 
The stacking profiles and associated energy costs are identical if either the strained (Fig. \ref{fig:energetics}(a)) or unstrained layer (Fig. \ref{fig:energetics}(b)) are relaxed, with the energy costs decreasing linearly for sharper interfaces (smaller $\alpha$).
In contrast, the strain energy cost is always higher if we chose to relax the initially unstrained layer.
In this case, large portions of both layers require strain values near $1\%$ in order to maintain the stacking order in the larger AB/BA regions found for smaller $\alpha$.
For both relaxation scenarios, and for all values of $\alpha$, the strain energy cost is greater than that of the uniformly-strained system (blue squares).
This is trivial if the unstrained layer is relaxed, as the cost of the additional strain introduced by the relaxation of this layer must be added to the cost of the uniformly strained layer.
When the strained layer is relaxed, the strain is redistributed throughout the layer such that the average strain in the layer is conserved. 
As the energy cost of strain increases superlinearly~\cite{georgoulea2022strain}, the additional energy cost of higher strain near the interfaces is not compensated by the reduced strain in the AB and BA domains.

Regardless of which layer is relaxed, the total energy curve is relatively flat over a wide range of interface smoothness values, until it increases sharply for $\alpha \lesssim 8$.
This is due to the significant strain energy cost for sharper interfaces, which becomes much larger than the energy savings due to more preferable stacking configurations.
There is a slight reduction in the total energy, compared to the uniformly strained case (black square symbol), if the strained (unstrained) layer is relaxed using $\alpha \gtrsim 8$ ($10$).
This is near the cut-off where a pseudogap, and associated topological interface states, emerge for smaller values of bias, as shown in Fig.~\ref{fig:alphas}(b). 
However, a reasonable energy window of order $\sim 100$ meV in which to observe topological states should occur for a larger interlayer bias of $\Delta = -400$ meV. 

Before examining interface states in more detail, it is worth considering additional factors which may affect strain redistribution in heterostrained bilayer graphene.
The relaxation model employed here is a simplification that allowed us to examine, in a general manner, the role of interface sharpness in such systems.
Substrate effects are neglected, except under the simplifying assumption that relaxation only occurs in one layer of the system.
The specific nature of the interaction between the substrate, or any tips or contacts used for either the application of strain or other measurements, may also effect the local distribution of strain in the system.
Similarly, we have not considered the role which out-of-plane deformations could play in reducing the total energy of heterostrained systems.
Indeed, a recent work shows that such deformations can affect flat band formation and local chemical reactivity in this type of system~\cite{schleder2023onedimensional}.
The local stacking energy will also be sensitive to modulations of the interlayer separation introduced by such deformations~\cite{gargiulo2017structural}.
Finally, we note that for simplicity we have only considered a single value of $\alpha$ within each structure so that the local strain at the AA and SP interfaces is the same. However, these interfaces have different local stacking energetics, relative shifts and widths, and so may be able to sustain different levels of local strain. 
We do not expect these additional considerations to dramatically alter the main trends in our results. 
The main effect is likely to be that these more complicated relaxed geometries could reduce the energy costs due to strain beyond the values predicted by our simple model, leading to an increased preference for sharper interfaces.
To conclude this section, we re-emphasise that uniform heterostrain and bias alone are not sufficient to create one-dimensional topological channels: an accompanying relaxation which sharpens the interfaces and expands the Bernal-stacked regions is also required.

\section{Interface states}
\label{sec:interfaces}

\begin{figure*}[]
    \center
    \includegraphics[width=.95\linewidth]{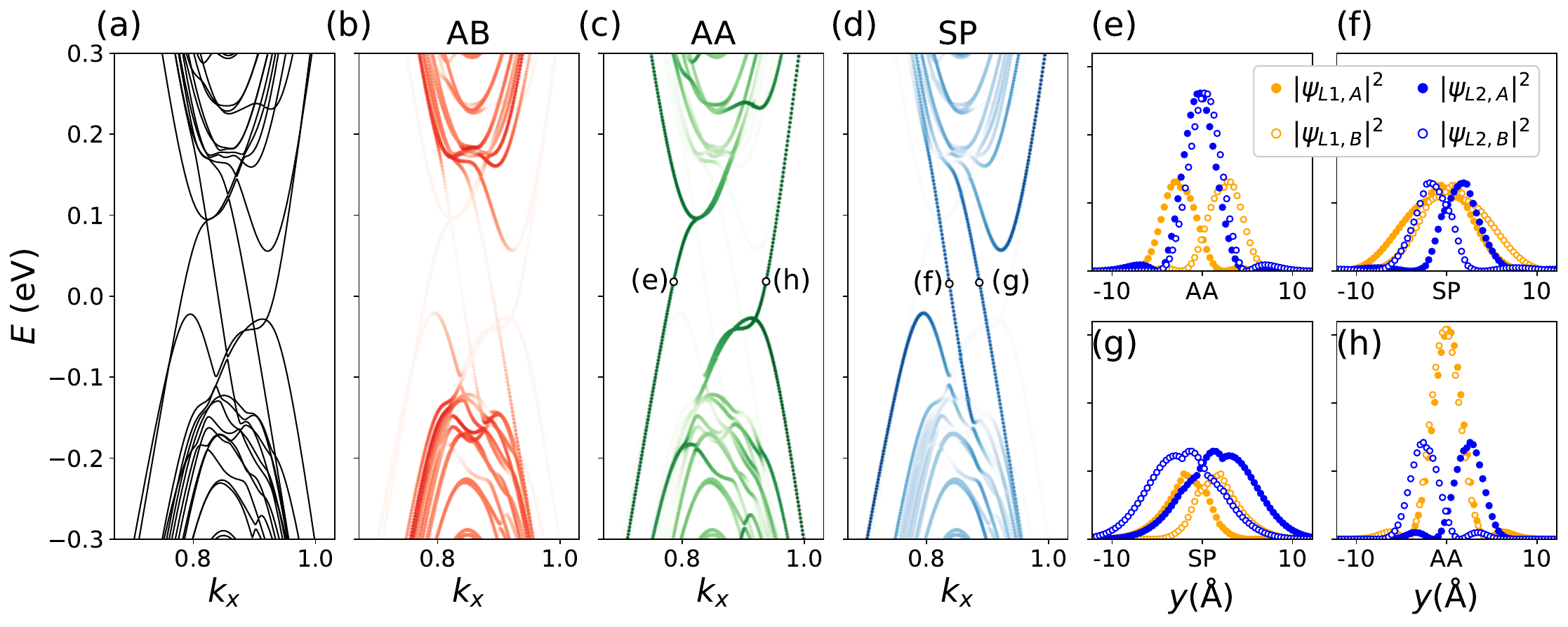}
    \caption{(a) Band structure of a heterostrained BLG system with an interlayer bias $\Delta = -400$ meV, where the interfaces are sharpened by a relaxation with $\alpha = 9$ in the strained layer. (b), (c), (d) show the band structure projected onto AB, AA and SP regions of the system, as in Figs. \ref{fig:bands} and \ref{fig:bands_bias}. Clear topological channels are formed at the $AA$ and $SP$ interfaces and are marked in panels (c) and (d). 
    The distributions of the these states around their respective interfaces are shown in panels (e)--(g). } 
    \label{fig:top_states}
\end{figure*}

Fig. ~\ref{fig:top_states}(a) shows the band structure near the K point of a heterostrained BLG system with the relaxation model applied to the strained layer, with $\alpha = 9$ and $\Delta = -400$ meV. 
From Figs. ~\ref{fig:alphas}(b) and ~\ref{fig:energetics}(a), these parameters are consistent with a sizeable pseudogap and reasonable energetic costs. 
A similar pseudogap could also be achieved for a lower bias, as shown for $\Delta=-200$ meV in Fig. \ref{fig:alphas}(a), but at the cost of a much higher energetic cost due to larger local strains near a sharper interface.
Conversely, achieving a similar pseudogap with a smoother interface would require a larger bias to be applied between the layers.
The band structure in Fig. ~\ref{fig:top_states}(a) shows a pseudogap spanned by four dispersive bands.
That these indeed correspond to interface-localised channels is made clear by the band projections in panels (b)--(d). 
The AB regions are gapped, whereas the AA- and SP- interfaces each host a pair of boundary modes in the K valley.
We note that the modes at a particular interface propagate in only a single direction for this valley, while the corresponding modes in the $K^\prime$ valley (not shown) propagate in the opposite direction. 
We further note that the $K$ valley modes from the two different interfaces have opposite propagation directions.
This is due to the opposite placements of the AB and BA regions relative to the two interfaces.
All these features are consistent with the formation of valley-protected topological modes at the interfaces due to a change in the valley Chern number between AB and BA regions in biased BLG~\cite{zhang2013valley}.

Finally, we note that there are subtle differences between modes at AA- or SP-interfaces.
Firstly, the energy window for observing these states is slightly different, due to the onset of higher-energy bands at different energies in the AA and SP regions.
We note that AA interface in our system is broader than the SP interface, as discussed in Sec. \ref{sec:simple}.
This does not seem to correspond to a wider interface state however.
Fig. ~\ref{fig:top_states}(e)--(h) show the projections of the corresponding states in panels (c) and (d) onto each atom in a $20$~\AA\ strip around the relevant interface.
A slightly broader spread is noted for the SP interface modes in (f), (g) compared to the AA  modes in panels (e), (h).
There is a more significant difference however in the layer (colour) and sublattice (symbol) distribution of these modes.
Each of the two AA modes (e, h) has largest weight exactly at the interface on one of the two layers, with both sublattices having very similar distributions in the dominant layer.
The other layer has a smaller, sublattice-split distribution with peaks to either side of the interface. 
In contrast, the SP modes (f, g) appear to have a more symmetric distribution between both layers and sublattices. 
We note that these trends are likely to be sensitive to the strain levels at the interfaces, and may not be present if, for example, relaxation results in different local strains at each interface.

\section{Conclusions}

Twisted bilayer graphene systems and the associated  Moir\'e superlattice have come under intense scrutiny recently due to the emergence of flat electronic bands, strongly correlated states, and unconventional superconductivity in this system. 
However, twisting is just one way to generate a Moir\'e superlattice. 
In this study we use tight-binding methods to study the formation and evolution of interface states in heterostrained bilayer graphene. 
Uniaxial heterostrain in this system gives rise to a one-dimensional Moir\'{e} pattern with two interface types between AB and BA regions, namely those with an SP-stacking and those with an AA-stacking.
Although uniform heterostrain creates domain boundaries between AB- and BA- regions, as in twisted bilayer graphene (tBLG), they are too wide to show clear signatures of topological interface channels. 
We demonstrate that in-plane relaxations, which maximise AB/BA-stacked domains and sharpen AA- and SP-stacked interfaces, are necessary for robust one-dimensional topologically protected channels to emerge.

By comparing the electronic band structure of a biased heterostrained bilayer to those of regularly-stacked systems, we identify bands that are highly-localised in AA- or SP-stacked interfaces. 
These bands emerge in the pseudogap which opens in the AB-stacked regions with the application of an interlayer bias. 
We show that the size of the pseudogap and the emergence of interface states depends on how sharp the interface is as well as the magnitude of applied bias. 
We show that when the AB regions are gapped, the AA- and SP- interfaces each host a pair of boundary modes in the K valley propagating unidirectionally. 
The corresponding modes in the K$^\prime$ valley propagate in the opposite direction. 
This is consistent with the formation of valley-protected topological modes at the interfaces due to a change in the valley Chern number between AB and BA regions in biased BLG.
Finally, unlike tBLG where only the SP-like interfaces contribute to the network, the Moir\'{e} system generated by the application of a uniaxial strain has two distinct interface states, namely those at the AA- and SP-stacked interfaces.
These interface states at the AA- and SP-interfaces are not identical, with differences in the layer and sublattice distribution of these modes.
We expect the exact nature of the distribution of these modes to depend sensitively on the strain levels at these interfaces. 

So far we have considered a pure uniaxial strain which creates a strictly one-dimensional modulation of the stacking order. 
Allowing for a biaxial strain, or a Poisson compression of the strained layer, produces a two-dimensional Moir\'e pattern~\cite{parhizkar2022strained}. 
We are now investigating how this will modify the emergent network of interface states. 
We expect that the channels associated with the AA-stacked interface will no longer contribute to the network, as in tBLG, while those associated with the SP-regions will remain. 
However, their configuration is likely to be different to those in tBLG. 

Heterostrained, untwisted bilayer graphene, as considered here, is a very promising platform from which to tune the presence and distribution of topological channels. This system could potentially exhibit phenomena akin to those in tBLG while circumventing some of the limitations associated with achieving precise twist angle control~\cite{kapfer2022programming}.

\bibliography{references1}

\begin{thebibliography}{70}%
\makeatletter
\providecommand \@ifxundefined [1]{%
 \@ifx{#1\undefined}
}%
\providecommand \@ifnum [1]{%
 \ifnum #1\expandafter \@firstoftwo
 \else \expandafter \@secondoftwo
 \fi
}%
\providecommand \@ifx [1]{%
 \ifx #1\expandafter \@firstoftwo
 \else \expandafter \@secondoftwo
 \fi
}%
\providecommand \natexlab [1]{#1}%
\providecommand \enquote  [1]{``#1''}%
\providecommand \bibnamefont  [1]{#1}%
\providecommand \bibfnamefont [1]{#1}%
\providecommand \citenamefont [1]{#1}%
\providecommand \href@noop [0]{\@secondoftwo}%
\providecommand \href [0]{\begingroup \@sanitize@url \@href}%
\providecommand \@href[1]{\@@startlink{#1}\@@href}%
\providecommand \@@href[1]{\endgroup#1\@@endlink}%
\providecommand \@sanitize@url [0]{\catcode `\\12\catcode `\$12\catcode
  `\&12\catcode `\#12\catcode `\^12\catcode `\_12\catcode `\%12\relax}%
\providecommand \@@startlink[1]{}%
\providecommand \@@endlink[0]{}%
\providecommand \url  [0]{\begingroup\@sanitize@url \@url }%
\providecommand \@url [1]{\endgroup\@href {#1}{\urlprefix }}%
\providecommand \urlprefix  [0]{URL }%
\providecommand \Eprint [0]{\href }%
\providecommand \doibase [0]{http://dx.doi.org/}%
\providecommand \selectlanguage [0]{\@gobble}%
\providecommand \bibinfo  [0]{\@secondoftwo}%
\providecommand \bibfield  [0]{\@secondoftwo}%
\providecommand \translation [1]{[#1]}%
\providecommand \BibitemOpen [0]{}%
\providecommand \bibitemStop [0]{}%
\providecommand \bibitemNoStop [0]{.\EOS\space}%
\providecommand \EOS [0]{\spacefactor3000\relax}%
\providecommand \BibitemShut  [1]{\csname bibitem#1\endcsname}%
\let\auto@bib@innerbib\@empty
\bibitem [{\citenamefont {Novoselov}\ \emph {et~al.}(2004)\citenamefont
  {Novoselov}, \citenamefont {Geim}, \citenamefont {Morozov}, \citenamefont
  {Jiang}, \citenamefont {Zhang}, \citenamefont {Dubonos}, \citenamefont
  {Grigorieva},\ and\ \citenamefont {Firsov}}]{novoselov2004electric}%
  \BibitemOpen
  \bibfield  {author} {\bibinfo {author} {\bibfnamefont {K.~S.}\ \bibnamefont
  {Novoselov}}, \bibinfo {author} {\bibfnamefont {A.~K.}\ \bibnamefont {Geim}},
  \bibinfo {author} {\bibfnamefont {S.~V.}\ \bibnamefont {Morozov}}, \bibinfo
  {author} {\bibfnamefont {D.-E.}\ \bibnamefont {Jiang}}, \bibinfo {author}
  {\bibfnamefont {Y.}~\bibnamefont {Zhang}}, \bibinfo {author} {\bibfnamefont
  {S.~V.}\ \bibnamefont {Dubonos}}, \bibinfo {author} {\bibfnamefont {I.~V.}\
  \bibnamefont {Grigorieva}}, \ and\ \bibinfo {author} {\bibfnamefont {A.~A.}\
  \bibnamefont {Firsov}},\ }\href@noop {} {\bibfield  {journal} {\bibinfo
  {journal} {Science}\ }\textbf {\bibinfo {volume} {306}},\ \bibinfo {pages}
  {666} (\bibinfo {year} {2004})}\BibitemShut {NoStop}%
\bibitem [{\citenamefont {Novoselov}\ \emph {et~al.}(2005)\citenamefont
  {Novoselov}, \citenamefont {Jiang}, \citenamefont {Schedin}, \citenamefont
  {Booth}, \citenamefont {Khotkevich}, \citenamefont {Morozov},\ and\
  \citenamefont {Geim}}]{novoselov2005two}%
  \BibitemOpen
  \bibfield  {author} {\bibinfo {author} {\bibfnamefont {K.~S.}\ \bibnamefont
  {Novoselov}}, \bibinfo {author} {\bibfnamefont {D.}~\bibnamefont {Jiang}},
  \bibinfo {author} {\bibfnamefont {F.}~\bibnamefont {Schedin}}, \bibinfo
  {author} {\bibfnamefont {T.}~\bibnamefont {Booth}}, \bibinfo {author}
  {\bibfnamefont {V.}~\bibnamefont {Khotkevich}}, \bibinfo {author}
  {\bibfnamefont {S.}~\bibnamefont {Morozov}}, \ and\ \bibinfo {author}
  {\bibfnamefont {A.~K.}\ \bibnamefont {Geim}},\ }\href@noop {} {\bibfield
  {journal} {\bibinfo  {journal} {Proceedings of the National Academy of
  Sciences}\ }\textbf {\bibinfo {volume} {102}},\ \bibinfo {pages} {10451}
  (\bibinfo {year} {2005})}\BibitemShut {NoStop}%
\bibitem [{\citenamefont {Katsnelson}(2007)}]{katsnelson2007graphene}%
  \BibitemOpen
  \bibfield  {author} {\bibinfo {author} {\bibfnamefont {M.~I.}\ \bibnamefont
  {Katsnelson}},\ }\href@noop {} {\bibfield  {journal} {\bibinfo  {journal}
  {Materials today}\ }\textbf {\bibinfo {volume} {10}},\ \bibinfo {pages} {20}
  (\bibinfo {year} {2007})}\BibitemShut {NoStop}%
\bibitem [{\citenamefont {Lee}\ \emph {et~al.}(2008)\citenamefont {Lee},
  \citenamefont {Wei}, \citenamefont {Kysar},\ and\ \citenamefont
  {Hone}}]{lee2008measurement}%
  \BibitemOpen
  \bibfield  {author} {\bibinfo {author} {\bibfnamefont {C.}~\bibnamefont
  {Lee}}, \bibinfo {author} {\bibfnamefont {X.}~\bibnamefont {Wei}}, \bibinfo
  {author} {\bibfnamefont {J.~W.}\ \bibnamefont {Kysar}}, \ and\ \bibinfo
  {author} {\bibfnamefont {J.}~\bibnamefont {Hone}},\ }\href@noop {} {\bibfield
   {journal} {\bibinfo  {journal} {science}\ }\textbf {\bibinfo {volume}
  {321}},\ \bibinfo {pages} {385} (\bibinfo {year} {2008})}\BibitemShut
  {NoStop}%
\bibitem [{\citenamefont {Wallace}(1947)}]{wallace1947band}%
  \BibitemOpen
  \bibfield  {author} {\bibinfo {author} {\bibfnamefont {P.~R.}\ \bibnamefont
  {Wallace}},\ }\href@noop {} {\bibfield  {journal} {\bibinfo  {journal}
  {Physical review}\ }\textbf {\bibinfo {volume} {71}},\ \bibinfo {pages} {622}
  (\bibinfo {year} {1947})}\BibitemShut {NoStop}%
\bibitem [{\citenamefont {Castro~Neto}\ \emph {et~al.}(2009)\citenamefont
  {Castro~Neto}, \citenamefont {Guinea}, \citenamefont {Peres}, \citenamefont
  {Novoselov},\ and\ \citenamefont {Geim}}]{castroneto:graphenereview}%
  \BibitemOpen
  \bibfield  {author} {\bibinfo {author} {\bibfnamefont {A.~H.}\ \bibnamefont
  {Castro~Neto}}, \bibinfo {author} {\bibfnamefont {F.}~\bibnamefont {Guinea}},
  \bibinfo {author} {\bibfnamefont {N.~M.~R.}\ \bibnamefont {Peres}}, \bibinfo
  {author} {\bibfnamefont {K.~S.}\ \bibnamefont {Novoselov}}, \ and\ \bibinfo
  {author} {\bibfnamefont {A.~K.}\ \bibnamefont {Geim}},\ }\href {\doibase
  10.1103/RevModPhys.81.109} {\bibfield  {journal} {\bibinfo  {journal} {Rev.
  Mod. Phys.}\ }\textbf {\bibinfo {volume} {81}},\ \bibinfo {pages} {109}
  (\bibinfo {year} {2009})}\BibitemShut {NoStop}%
\bibitem [{\citenamefont {Katsnelson}\ \emph {et~al.}(2006)\citenamefont
  {Katsnelson}, \citenamefont {Novoselov},\ and\ \citenamefont
  {Geim}}]{katsnelson2006chiral}%
  \BibitemOpen
  \bibfield  {author} {\bibinfo {author} {\bibfnamefont {M.~I.}\ \bibnamefont
  {Katsnelson}}, \bibinfo {author} {\bibfnamefont {K.~S.}\ \bibnamefont
  {Novoselov}}, \ and\ \bibinfo {author} {\bibfnamefont {A.~K.}\ \bibnamefont
  {Geim}},\ }\href@noop {} {\bibfield  {journal} {\bibinfo  {journal} {Nature
  physics}\ }\textbf {\bibinfo {volume} {2}},\ \bibinfo {pages} {620} (\bibinfo
  {year} {2006})}\BibitemShut {NoStop}%
\bibitem [{\citenamefont {Novoselov}\ \emph {et~al.}(2012)\citenamefont
  {Novoselov}, \citenamefont {Fal}, \citenamefont {Colombo}, \citenamefont
  {Gellert}, \citenamefont {Schwab}, \citenamefont {Kim} \emph
  {et~al.}}]{novoselov2012roadmap}%
  \BibitemOpen
  \bibfield  {author} {\bibinfo {author} {\bibfnamefont {K.~S.}\ \bibnamefont
  {Novoselov}}, \bibinfo {author} {\bibfnamefont {V.}~\bibnamefont {Fal}},
  \bibinfo {author} {\bibfnamefont {L.}~\bibnamefont {Colombo}}, \bibinfo
  {author} {\bibfnamefont {P.}~\bibnamefont {Gellert}}, \bibinfo {author}
  {\bibfnamefont {M.}~\bibnamefont {Schwab}}, \bibinfo {author} {\bibfnamefont
  {K.}~\bibnamefont {Kim}},  \emph {et~al.},\ }\href@noop {} {\bibfield
  {journal} {\bibinfo  {journal} {Nature}\ }\textbf {\bibinfo {volume} {490}},\
  \bibinfo {pages} {192} (\bibinfo {year} {2012})}\BibitemShut {NoStop}%
\bibitem [{\citenamefont {Bae}\ \emph {et~al.}(2010)\citenamefont {Bae},
  \citenamefont {Kim}, \citenamefont {Lee}, \citenamefont {Xu}, \citenamefont
  {Park}, \citenamefont {Zheng}, \citenamefont {Balakrishnan}, \citenamefont
  {Lei}, \citenamefont {Kim}, \citenamefont {Song} \emph
  {et~al.}}]{bae2010roll}%
  \BibitemOpen
  \bibfield  {author} {\bibinfo {author} {\bibfnamefont {S.}~\bibnamefont
  {Bae}}, \bibinfo {author} {\bibfnamefont {H.}~\bibnamefont {Kim}}, \bibinfo
  {author} {\bibfnamefont {Y.}~\bibnamefont {Lee}}, \bibinfo {author}
  {\bibfnamefont {X.}~\bibnamefont {Xu}}, \bibinfo {author} {\bibfnamefont
  {J.-S.}\ \bibnamefont {Park}}, \bibinfo {author} {\bibfnamefont
  {Y.}~\bibnamefont {Zheng}}, \bibinfo {author} {\bibfnamefont
  {J.}~\bibnamefont {Balakrishnan}}, \bibinfo {author} {\bibfnamefont
  {T.}~\bibnamefont {Lei}}, \bibinfo {author} {\bibfnamefont {H.~R.}\
  \bibnamefont {Kim}}, \bibinfo {author} {\bibfnamefont {Y.~I.}\ \bibnamefont
  {Song}},  \emph {et~al.},\ }\href@noop {} {\bibfield  {journal} {\bibinfo
  {journal} {Nature nanotechnology}\ }\textbf {\bibinfo {volume} {5}},\
  \bibinfo {pages} {574} (\bibinfo {year} {2010})}\BibitemShut {NoStop}%
\bibitem [{\citenamefont {Han}\ \emph {et~al.}(2012)\citenamefont {Han},
  \citenamefont {Lee}, \citenamefont {Choi}, \citenamefont {Woo}, \citenamefont
  {Bae}, \citenamefont {Hong}, \citenamefont {Ahn},\ and\ \citenamefont
  {Lee}}]{han2012extremely}%
  \BibitemOpen
  \bibfield  {author} {\bibinfo {author} {\bibfnamefont {T.-H.}\ \bibnamefont
  {Han}}, \bibinfo {author} {\bibfnamefont {Y.}~\bibnamefont {Lee}}, \bibinfo
  {author} {\bibfnamefont {M.-R.}\ \bibnamefont {Choi}}, \bibinfo {author}
  {\bibfnamefont {S.-H.}\ \bibnamefont {Woo}}, \bibinfo {author} {\bibfnamefont
  {S.-H.}\ \bibnamefont {Bae}}, \bibinfo {author} {\bibfnamefont {B.~H.}\
  \bibnamefont {Hong}}, \bibinfo {author} {\bibfnamefont {J.-H.}\ \bibnamefont
  {Ahn}}, \ and\ \bibinfo {author} {\bibfnamefont {T.-W.}\ \bibnamefont
  {Lee}},\ }\href@noop {} {\bibfield  {journal} {\bibinfo  {journal} {Nature
  Photonics}\ }\textbf {\bibinfo {volume} {6}},\ \bibinfo {pages} {105}
  (\bibinfo {year} {2012})}\BibitemShut {NoStop}%
\bibitem [{\citenamefont {Geim}\ and\ \citenamefont
  {Grigorieva}(2013)}]{geim2013van}%
  \BibitemOpen
  \bibfield  {author} {\bibinfo {author} {\bibfnamefont {A.~K.}\ \bibnamefont
  {Geim}}\ and\ \bibinfo {author} {\bibfnamefont {I.~V.}\ \bibnamefont
  {Grigorieva}},\ }\href@noop {} {\bibfield  {journal} {\bibinfo  {journal}
  {Nature}\ }\textbf {\bibinfo {volume} {499}},\ \bibinfo {pages} {419}
  (\bibinfo {year} {2013})}\BibitemShut {NoStop}%
\bibitem [{\citenamefont {McCann}\ and\ \citenamefont
  {Koshino}(2013)}]{mccann2013electronic}%
  \BibitemOpen
  \bibfield  {author} {\bibinfo {author} {\bibfnamefont {E.}~\bibnamefont
  {McCann}}\ and\ \bibinfo {author} {\bibfnamefont {M.}~\bibnamefont
  {Koshino}},\ }\href@noop {} {\bibfield  {journal} {\bibinfo  {journal}
  {Reports on Progress in physics}\ }\textbf {\bibinfo {volume} {76}},\
  \bibinfo {pages} {056503} (\bibinfo {year} {2013})}\BibitemShut {NoStop}%
\bibitem [{\citenamefont {Azadmanjiri}\ \emph {et~al.}(2020)\citenamefont
  {Azadmanjiri}, \citenamefont {Srivastava}, \citenamefont {Kumar},
  \citenamefont {Sofer}, \citenamefont {Min},\ and\ \citenamefont
  {Gong}}]{azadmanjiri2020graphene}%
  \BibitemOpen
  \bibfield  {author} {\bibinfo {author} {\bibfnamefont {J.}~\bibnamefont
  {Azadmanjiri}}, \bibinfo {author} {\bibfnamefont {V.~K.}\ \bibnamefont
  {Srivastava}}, \bibinfo {author} {\bibfnamefont {P.}~\bibnamefont {Kumar}},
  \bibinfo {author} {\bibfnamefont {Z.}~\bibnamefont {Sofer}}, \bibinfo
  {author} {\bibfnamefont {J.}~\bibnamefont {Min}}, \ and\ \bibinfo {author}
  {\bibfnamefont {J.}~\bibnamefont {Gong}},\ }\href@noop {} {\bibfield
  {journal} {\bibinfo  {journal} {Applied Materials Today}\ }\textbf {\bibinfo
  {volume} {19}},\ \bibinfo {pages} {100600} (\bibinfo {year}
  {2020})}\BibitemShut {NoStop}%
\bibitem [{\citenamefont {Rozhkov}\ \emph {et~al.}(2016)\citenamefont
  {Rozhkov}, \citenamefont {Sboychakov}, \citenamefont {Rakhmanov},\ and\
  \citenamefont {Nori}}]{rozhkov2016electronic}%
  \BibitemOpen
  \bibfield  {author} {\bibinfo {author} {\bibfnamefont {A.~V.}\ \bibnamefont
  {Rozhkov}}, \bibinfo {author} {\bibfnamefont {A.}~\bibnamefont {Sboychakov}},
  \bibinfo {author} {\bibfnamefont {A.}~\bibnamefont {Rakhmanov}}, \ and\
  \bibinfo {author} {\bibfnamefont {F.}~\bibnamefont {Nori}},\ }\href@noop {}
  {\bibfield  {journal} {\bibinfo  {journal} {Physics Reports}\ }\textbf
  {\bibinfo {volume} {648}},\ \bibinfo {pages} {1} (\bibinfo {year}
  {2016})}\BibitemShut {NoStop}%
\bibitem [{\citenamefont {Ho}\ \emph {et~al.}(2006)\citenamefont {Ho},
  \citenamefont {Lu}, \citenamefont {Hwang}, \citenamefont {Chang},\ and\
  \citenamefont {Lin}}]{ho2006coulomb}%
  \BibitemOpen
  \bibfield  {author} {\bibinfo {author} {\bibfnamefont {J.-H.}\ \bibnamefont
  {Ho}}, \bibinfo {author} {\bibfnamefont {C.}~\bibnamefont {Lu}}, \bibinfo
  {author} {\bibfnamefont {C.}~\bibnamefont {Hwang}}, \bibinfo {author}
  {\bibfnamefont {C.}~\bibnamefont {Chang}}, \ and\ \bibinfo {author}
  {\bibfnamefont {M.-F.}\ \bibnamefont {Lin}},\ }\href@noop {} {\bibfield
  {journal} {\bibinfo  {journal} {Physical Review B}\ }\textbf {\bibinfo
  {volume} {74}},\ \bibinfo {pages} {085406} (\bibinfo {year}
  {2006})}\BibitemShut {NoStop}%
\bibitem [{\citenamefont {Park}\ \emph {et~al.}(2015)\citenamefont {Park},
  \citenamefont {Ryou}, \citenamefont {Hong}, \citenamefont {Sumpter},
  \citenamefont {Kim},\ and\ \citenamefont {Yoon}}]{park2015electronic}%
  \BibitemOpen
  \bibfield  {author} {\bibinfo {author} {\bibfnamefont {C.}~\bibnamefont
  {Park}}, \bibinfo {author} {\bibfnamefont {J.}~\bibnamefont {Ryou}}, \bibinfo
  {author} {\bibfnamefont {S.}~\bibnamefont {Hong}}, \bibinfo {author}
  {\bibfnamefont {B.~G.}\ \bibnamefont {Sumpter}}, \bibinfo {author}
  {\bibfnamefont {G.}~\bibnamefont {Kim}}, \ and\ \bibinfo {author}
  {\bibfnamefont {M.}~\bibnamefont {Yoon}},\ }\href@noop {} {\bibfield
  {journal} {\bibinfo  {journal} {Physical review letters}\ }\textbf {\bibinfo
  {volume} {115}},\ \bibinfo {pages} {015502} (\bibinfo {year}
  {2015})}\BibitemShut {NoStop}%
\bibitem [{\citenamefont {Castro}\ \emph {et~al.}(2007)\citenamefont {Castro},
  \citenamefont {Novoselov}, \citenamefont {Morozov}, \citenamefont {Peres},
  \citenamefont {Dos~Santos}, \citenamefont {Nilsson}, \citenamefont {Guinea},
  \citenamefont {Geim},\ and\ \citenamefont {Castro~Neto}}]{castro2007biased}%
  \BibitemOpen
  \bibfield  {author} {\bibinfo {author} {\bibfnamefont {E.~V.}\ \bibnamefont
  {Castro}}, \bibinfo {author} {\bibfnamefont {K.}~\bibnamefont {Novoselov}},
  \bibinfo {author} {\bibfnamefont {S.}~\bibnamefont {Morozov}}, \bibinfo
  {author} {\bibfnamefont {N.}~\bibnamefont {Peres}}, \bibinfo {author}
  {\bibfnamefont {J.~L.}\ \bibnamefont {Dos~Santos}}, \bibinfo {author}
  {\bibfnamefont {J.}~\bibnamefont {Nilsson}}, \bibinfo {author} {\bibfnamefont
  {F.}~\bibnamefont {Guinea}}, \bibinfo {author} {\bibfnamefont
  {A.}~\bibnamefont {Geim}}, \ and\ \bibinfo {author} {\bibfnamefont
  {A.}~\bibnamefont {Castro~Neto}},\ }\href@noop {} {\bibfield  {journal}
  {\bibinfo  {journal} {Physical review letters}\ }\textbf {\bibinfo {volume}
  {99}},\ \bibinfo {pages} {216802} (\bibinfo {year} {2007})}\BibitemShut
  {NoStop}%
\bibitem [{\citenamefont {Silva}\ \emph {et~al.}(2020)\citenamefont {Silva},
  \citenamefont {Santos}, \citenamefont {Skelton}, \citenamefont {Yang},
  \citenamefont {Santos}, \citenamefont {Parker},\ and\ \citenamefont
  {Walsh}}]{silva2020electronic}%
  \BibitemOpen
  \bibfield  {author} {\bibinfo {author} {\bibfnamefont {E.}~\bibnamefont
  {Silva}}, \bibinfo {author} {\bibfnamefont {M.}~\bibnamefont {Santos}},
  \bibinfo {author} {\bibfnamefont {J.}~\bibnamefont {Skelton}}, \bibinfo
  {author} {\bibfnamefont {T.}~\bibnamefont {Yang}}, \bibinfo {author}
  {\bibfnamefont {T.}~\bibnamefont {Santos}}, \bibinfo {author} {\bibfnamefont
  {S.}~\bibnamefont {Parker}}, \ and\ \bibinfo {author} {\bibfnamefont
  {A.}~\bibnamefont {Walsh}},\ }\href@noop {} {\bibfield  {journal} {\bibinfo
  {journal} {Materials Today: Proceedings}\ }\textbf {\bibinfo {volume} {20}},\
  \bibinfo {pages} {373} (\bibinfo {year} {2020})}\BibitemShut {NoStop}%
\bibitem [{\citenamefont {Zhang}\ \emph {et~al.}(2013)\citenamefont {Zhang},
  \citenamefont {MacDonald},\ and\ \citenamefont {Mele}}]{zhang2013valley}%
  \BibitemOpen
  \bibfield  {author} {\bibinfo {author} {\bibfnamefont {F.}~\bibnamefont
  {Zhang}}, \bibinfo {author} {\bibfnamefont {A.~H.}\ \bibnamefont
  {MacDonald}}, \ and\ \bibinfo {author} {\bibfnamefont {E.~J.}\ \bibnamefont
  {Mele}},\ }\href@noop {} {\bibfield  {journal} {\bibinfo  {journal}
  {Proceedings of the National Academy of Sciences}\ }\textbf {\bibinfo
  {volume} {110}},\ \bibinfo {pages} {10546} (\bibinfo {year}
  {2013})}\BibitemShut {NoStop}%
\bibitem [{\citenamefont {San-Jose}\ and\ \citenamefont
  {Prada}(2013)}]{san2013helical}%
  \BibitemOpen
  \bibfield  {author} {\bibinfo {author} {\bibfnamefont {P.}~\bibnamefont
  {San-Jose}}\ and\ \bibinfo {author} {\bibfnamefont {E.}~\bibnamefont
  {Prada}},\ }\href@noop {} {\bibfield  {journal} {\bibinfo  {journal}
  {Physical Review B}\ }\textbf {\bibinfo {volume} {88}},\ \bibinfo {pages}
  {121408} (\bibinfo {year} {2013})}\BibitemShut {NoStop}%
\bibitem [{\citenamefont {Vaezi}\ \emph {et~al.}(2013)\citenamefont {Vaezi},
  \citenamefont {Liang}, \citenamefont {Ngai}, \citenamefont {Yang},\ and\
  \citenamefont {Kim}}]{vaezi2013topological}%
  \BibitemOpen
  \bibfield  {author} {\bibinfo {author} {\bibfnamefont {A.}~\bibnamefont
  {Vaezi}}, \bibinfo {author} {\bibfnamefont {Y.}~\bibnamefont {Liang}},
  \bibinfo {author} {\bibfnamefont {D.~H.}\ \bibnamefont {Ngai}}, \bibinfo
  {author} {\bibfnamefont {L.}~\bibnamefont {Yang}}, \ and\ \bibinfo {author}
  {\bibfnamefont {E.-A.}\ \bibnamefont {Kim}},\ }\href@noop {} {\bibfield
  {journal} {\bibinfo  {journal} {Physical Review X}\ }\textbf {\bibinfo
  {volume} {3}},\ \bibinfo {pages} {021018} (\bibinfo {year}
  {2013})}\BibitemShut {NoStop}%
\bibitem [{\citenamefont {Ju}\ \emph {et~al.}(2015)\citenamefont {Ju},
  \citenamefont {Shi}, \citenamefont {Nair}, \citenamefont {Lv}, \citenamefont
  {Jin}, \citenamefont {Velasco}, \citenamefont {Ojeda-Aristizabal},
  \citenamefont {Bechtel}, \citenamefont {Martin}, \citenamefont {Zettl} \emph
  {et~al.}}]{ju2015topological}%
  \BibitemOpen
  \bibfield  {author} {\bibinfo {author} {\bibfnamefont {L.}~\bibnamefont
  {Ju}}, \bibinfo {author} {\bibfnamefont {Z.}~\bibnamefont {Shi}}, \bibinfo
  {author} {\bibfnamefont {N.}~\bibnamefont {Nair}}, \bibinfo {author}
  {\bibfnamefont {Y.}~\bibnamefont {Lv}}, \bibinfo {author} {\bibfnamefont
  {C.}~\bibnamefont {Jin}}, \bibinfo {author} {\bibfnamefont {J.}~\bibnamefont
  {Velasco}}, \bibinfo {author} {\bibfnamefont {C.}~\bibnamefont
  {Ojeda-Aristizabal}}, \bibinfo {author} {\bibfnamefont {H.~A.}\ \bibnamefont
  {Bechtel}}, \bibinfo {author} {\bibfnamefont {M.~C.}\ \bibnamefont {Martin}},
  \bibinfo {author} {\bibfnamefont {A.}~\bibnamefont {Zettl}},  \emph
  {et~al.},\ }\href@noop {} {\bibfield  {journal} {\bibinfo  {journal}
  {Nature}\ }\textbf {\bibinfo {volume} {520}},\ \bibinfo {pages} {650}
  (\bibinfo {year} {2015})}\BibitemShut {NoStop}%
\bibitem [{\citenamefont {Yasaei}\ \emph {et~al.}(2014)\citenamefont {Yasaei},
  \citenamefont {Kumar}, \citenamefont {Hantehzadeh}, \citenamefont {Kayyalha},
  \citenamefont {Baskin}, \citenamefont {Repnin}, \citenamefont {Wang},
  \citenamefont {Klie}, \citenamefont {Chen}, \citenamefont {Kr{\'a}l} \emph
  {et~al.}}]{yasaei2014chemical}%
  \BibitemOpen
  \bibfield  {author} {\bibinfo {author} {\bibfnamefont {P.}~\bibnamefont
  {Yasaei}}, \bibinfo {author} {\bibfnamefont {B.}~\bibnamefont {Kumar}},
  \bibinfo {author} {\bibfnamefont {R.}~\bibnamefont {Hantehzadeh}}, \bibinfo
  {author} {\bibfnamefont {M.}~\bibnamefont {Kayyalha}}, \bibinfo {author}
  {\bibfnamefont {A.}~\bibnamefont {Baskin}}, \bibinfo {author} {\bibfnamefont
  {N.}~\bibnamefont {Repnin}}, \bibinfo {author} {\bibfnamefont
  {C.}~\bibnamefont {Wang}}, \bibinfo {author} {\bibfnamefont {R.~F.}\
  \bibnamefont {Klie}}, \bibinfo {author} {\bibfnamefont {Y.~P.}\ \bibnamefont
  {Chen}}, \bibinfo {author} {\bibfnamefont {P.}~\bibnamefont {Kr{\'a}l}},
  \emph {et~al.},\ }\href@noop {} {\bibfield  {journal} {\bibinfo  {journal}
  {Nature communications}\ }\textbf {\bibinfo {volume} {5}},\ \bibinfo {pages}
  {4911} (\bibinfo {year} {2014})}\BibitemShut {NoStop}%
\bibitem [{\citenamefont {Jask{\'o}lski}\ \emph {et~al.}(2016)\citenamefont
  {Jask{\'o}lski}, \citenamefont {Pelc}, \citenamefont {Chico},\ and\
  \citenamefont {Ayuela}}]{jaskolski2016existence}%
  \BibitemOpen
  \bibfield  {author} {\bibinfo {author} {\bibfnamefont {W.}~\bibnamefont
  {Jask{\'o}lski}}, \bibinfo {author} {\bibfnamefont {M.}~\bibnamefont {Pelc}},
  \bibinfo {author} {\bibfnamefont {L.}~\bibnamefont {Chico}}, \ and\ \bibinfo
  {author} {\bibfnamefont {A.}~\bibnamefont {Ayuela}},\ }\href@noop {}
  {\bibfield  {journal} {\bibinfo  {journal} {Nanoscale}\ }\textbf {\bibinfo
  {volume} {8}},\ \bibinfo {pages} {6079} (\bibinfo {year} {2016})}\BibitemShut
  {NoStop}%
\bibitem [{\citenamefont {Oostinga}\ \emph {et~al.}(2008)\citenamefont
  {Oostinga}, \citenamefont {Heersche}, \citenamefont {Liu}, \citenamefont
  {Morpurgo},\ and\ \citenamefont {Vandersypen}}]{oostinga2008gate}%
  \BibitemOpen
  \bibfield  {author} {\bibinfo {author} {\bibfnamefont {J.~B.}\ \bibnamefont
  {Oostinga}}, \bibinfo {author} {\bibfnamefont {H.~B.}\ \bibnamefont
  {Heersche}}, \bibinfo {author} {\bibfnamefont {X.}~\bibnamefont {Liu}},
  \bibinfo {author} {\bibfnamefont {A.~F.}\ \bibnamefont {Morpurgo}}, \ and\
  \bibinfo {author} {\bibfnamefont {L.~M.}\ \bibnamefont {Vandersypen}},\
  }\href@noop {} {\bibfield  {journal} {\bibinfo  {journal} {Nature materials}\
  }\textbf {\bibinfo {volume} {7}},\ \bibinfo {pages} {151} (\bibinfo {year}
  {2008})}\BibitemShut {NoStop}%
\bibitem [{\citenamefont {Zhang}\ \emph {et~al.}(2009)\citenamefont {Zhang},
  \citenamefont {Tang}, \citenamefont {Girit}, \citenamefont {Hao},
  \citenamefont {Martin}, \citenamefont {Zettl}, \citenamefont {Crommie},
  \citenamefont {Shen},\ and\ \citenamefont {Wang}}]{zhang2009direct}%
  \BibitemOpen
  \bibfield  {author} {\bibinfo {author} {\bibfnamefont {Y.}~\bibnamefont
  {Zhang}}, \bibinfo {author} {\bibfnamefont {T.-T.}\ \bibnamefont {Tang}},
  \bibinfo {author} {\bibfnamefont {C.}~\bibnamefont {Girit}}, \bibinfo
  {author} {\bibfnamefont {Z.}~\bibnamefont {Hao}}, \bibinfo {author}
  {\bibfnamefont {M.~C.}\ \bibnamefont {Martin}}, \bibinfo {author}
  {\bibfnamefont {A.}~\bibnamefont {Zettl}}, \bibinfo {author} {\bibfnamefont
  {M.~F.}\ \bibnamefont {Crommie}}, \bibinfo {author} {\bibfnamefont {Y.~R.}\
  \bibnamefont {Shen}}, \ and\ \bibinfo {author} {\bibfnamefont
  {F.}~\bibnamefont {Wang}},\ }\href@noop {} {\bibfield  {journal} {\bibinfo
  {journal} {Nature}\ }\textbf {\bibinfo {volume} {459}},\ \bibinfo {pages}
  {820} (\bibinfo {year} {2009})}\BibitemShut {NoStop}%
\bibitem [{\citenamefont {Li}\ \emph {et~al.}(2016)\citenamefont {Li},
  \citenamefont {Wang}, \citenamefont {McFaul}, \citenamefont {Zern},
  \citenamefont {Ren}, \citenamefont {Watanabe}, \citenamefont {Taniguchi},
  \citenamefont {Qiao},\ and\ \citenamefont {Zhu}}]{li2016gate}%
  \BibitemOpen
  \bibfield  {author} {\bibinfo {author} {\bibfnamefont {J.}~\bibnamefont
  {Li}}, \bibinfo {author} {\bibfnamefont {K.}~\bibnamefont {Wang}}, \bibinfo
  {author} {\bibfnamefont {K.~J.}\ \bibnamefont {McFaul}}, \bibinfo {author}
  {\bibfnamefont {Z.}~\bibnamefont {Zern}}, \bibinfo {author} {\bibfnamefont
  {Y.}~\bibnamefont {Ren}}, \bibinfo {author} {\bibfnamefont {K.}~\bibnamefont
  {Watanabe}}, \bibinfo {author} {\bibfnamefont {T.}~\bibnamefont {Taniguchi}},
  \bibinfo {author} {\bibfnamefont {Z.}~\bibnamefont {Qiao}}, \ and\ \bibinfo
  {author} {\bibfnamefont {J.}~\bibnamefont {Zhu}},\ }\href@noop {} {\bibfield
  {journal} {\bibinfo  {journal} {Nature nanotechnology}\ }\textbf {\bibinfo
  {volume} {11}},\ \bibinfo {pages} {1060} (\bibinfo {year}
  {2016})}\BibitemShut {NoStop}%
\bibitem [{\citenamefont {Rickhaus}\ \emph {et~al.}(2018)\citenamefont
  {Rickhaus}, \citenamefont {Wallbank}, \citenamefont {Slizovskiy},
  \citenamefont {Pisoni}, \citenamefont {Overweg}, \citenamefont {Lee},
  \citenamefont {Eich}, \citenamefont {Liu}, \citenamefont {Watanabe},
  \citenamefont {Taniguchi} \emph {et~al.}}]{rickhaus2018transport}%
  \BibitemOpen
  \bibfield  {author} {\bibinfo {author} {\bibfnamefont {P.}~\bibnamefont
  {Rickhaus}}, \bibinfo {author} {\bibfnamefont {J.}~\bibnamefont {Wallbank}},
  \bibinfo {author} {\bibfnamefont {S.}~\bibnamefont {Slizovskiy}}, \bibinfo
  {author} {\bibfnamefont {R.}~\bibnamefont {Pisoni}}, \bibinfo {author}
  {\bibfnamefont {H.}~\bibnamefont {Overweg}}, \bibinfo {author} {\bibfnamefont
  {Y.}~\bibnamefont {Lee}}, \bibinfo {author} {\bibfnamefont {M.}~\bibnamefont
  {Eich}}, \bibinfo {author} {\bibfnamefont {M.-H.}\ \bibnamefont {Liu}},
  \bibinfo {author} {\bibfnamefont {K.}~\bibnamefont {Watanabe}}, \bibinfo
  {author} {\bibfnamefont {T.}~\bibnamefont {Taniguchi}},  \emph {et~al.},\
  }\href@noop {} {\bibfield  {journal} {\bibinfo  {journal} {Nano letters}\
  }\textbf {\bibinfo {volume} {18}},\ \bibinfo {pages} {6725} (\bibinfo {year}
  {2018})}\BibitemShut {NoStop}%
\bibitem [{\citenamefont {Bistritzer}\ and\ \citenamefont
  {MacDonald}(2011)}]{bistritzer2011moire}%
  \BibitemOpen
  \bibfield  {author} {\bibinfo {author} {\bibfnamefont {R.}~\bibnamefont
  {Bistritzer}}\ and\ \bibinfo {author} {\bibfnamefont {A.~H.}\ \bibnamefont
  {MacDonald}},\ }\href@noop {} {\bibfield  {journal} {\bibinfo  {journal}
  {Proceedings of the National Academy of Sciences}\ }\textbf {\bibinfo
  {volume} {108}},\ \bibinfo {pages} {12233} (\bibinfo {year}
  {2011})}\BibitemShut {NoStop}%
\bibitem [{\citenamefont {Andrei}\ and\ \citenamefont
  {MacDonald}(2020)}]{andrei2020graphene}%
  \BibitemOpen
  \bibfield  {author} {\bibinfo {author} {\bibfnamefont {E.~Y.}\ \bibnamefont
  {Andrei}}\ and\ \bibinfo {author} {\bibfnamefont {A.~H.}\ \bibnamefont
  {MacDonald}},\ }\href@noop {} {\bibfield  {journal} {\bibinfo  {journal}
  {Nature materials}\ }\textbf {\bibinfo {volume} {19}},\ \bibinfo {pages}
  {1265} (\bibinfo {year} {2020})}\BibitemShut {NoStop}%
\bibitem [{\citenamefont {Efimkin}\ and\ \citenamefont
  {MacDonald}(2018)}]{efimkin2018helical}%
  \BibitemOpen
  \bibfield  {author} {\bibinfo {author} {\bibfnamefont {D.~K.}\ \bibnamefont
  {Efimkin}}\ and\ \bibinfo {author} {\bibfnamefont {A.~H.}\ \bibnamefont
  {MacDonald}},\ }\href {\doibase 10.1103/PhysRevB.98.035404} {\bibfield
  {journal} {\bibinfo  {journal} {Phys. Rev. B}\ }\textbf {\bibinfo {volume}
  {98}},\ \bibinfo {pages} {035404} (\bibinfo {year} {2018})}\BibitemShut
  {NoStop}%
\bibitem [{\citenamefont {Fleischmann}\ \emph {et~al.}(2019)\citenamefont
  {Fleischmann}, \citenamefont {Gupta}, \citenamefont {Wullschl{\"a}ger},
  \citenamefont {Theil}, \citenamefont {Weckbecker}, \citenamefont {Meded},
  \citenamefont {Sharma}, \citenamefont {Meyer},\ and\ \citenamefont
  {Shallcross}}]{fleischmann2019perfect}%
  \BibitemOpen
  \bibfield  {author} {\bibinfo {author} {\bibfnamefont {M.}~\bibnamefont
  {Fleischmann}}, \bibinfo {author} {\bibfnamefont {R.}~\bibnamefont {Gupta}},
  \bibinfo {author} {\bibfnamefont {F.}~\bibnamefont {Wullschl{\"a}ger}},
  \bibinfo {author} {\bibfnamefont {S.}~\bibnamefont {Theil}}, \bibinfo
  {author} {\bibfnamefont {D.}~\bibnamefont {Weckbecker}}, \bibinfo {author}
  {\bibfnamefont {V.}~\bibnamefont {Meded}}, \bibinfo {author} {\bibfnamefont
  {S.}~\bibnamefont {Sharma}}, \bibinfo {author} {\bibfnamefont
  {B.}~\bibnamefont {Meyer}}, \ and\ \bibinfo {author} {\bibfnamefont
  {S.}~\bibnamefont {Shallcross}},\ }\href@noop {} {\bibfield  {journal}
  {\bibinfo  {journal} {Nano Letters}\ }\textbf {\bibinfo {volume} {20}},\
  \bibinfo {pages} {971} (\bibinfo {year} {2019})}\BibitemShut {NoStop}%
\bibitem [{\citenamefont {Tsim}\ \emph {et~al.}(2020)\citenamefont {Tsim},
  \citenamefont {Nam},\ and\ \citenamefont {Koshino}}]{tsim2020perfect}%
  \BibitemOpen
  \bibfield  {author} {\bibinfo {author} {\bibfnamefont {B.}~\bibnamefont
  {Tsim}}, \bibinfo {author} {\bibfnamefont {N.~N.}\ \bibnamefont {Nam}}, \
  and\ \bibinfo {author} {\bibfnamefont {M.}~\bibnamefont {Koshino}},\
  }\href@noop {} {\bibfield  {journal} {\bibinfo  {journal} {Physical Review
  B}\ }\textbf {\bibinfo {volume} {101}},\ \bibinfo {pages} {125409} (\bibinfo
  {year} {2020})}\BibitemShut {NoStop}%
\bibitem [{\citenamefont {De~Beule}\ \emph {et~al.}(2021)\citenamefont
  {De~Beule}, \citenamefont {Dominguez},\ and\ \citenamefont
  {Recher}}]{de2021network}%
  \BibitemOpen
  \bibfield  {author} {\bibinfo {author} {\bibfnamefont {C.}~\bibnamefont
  {De~Beule}}, \bibinfo {author} {\bibfnamefont {F.}~\bibnamefont {Dominguez}},
  \ and\ \bibinfo {author} {\bibfnamefont {P.}~\bibnamefont {Recher}},\
  }\href@noop {} {\bibfield  {journal} {\bibinfo  {journal} {Physical Review
  B}\ }\textbf {\bibinfo {volume} {104}},\ \bibinfo {pages} {195410} (\bibinfo
  {year} {2021})}\BibitemShut {NoStop}%
\bibitem [{\citenamefont {Huang}\ \emph {et~al.}(2018)\citenamefont {Huang},
  \citenamefont {Kim}, \citenamefont {Efimkin}, \citenamefont {Lovorn},
  \citenamefont {Taniguchi}, \citenamefont {Watanabe}, \citenamefont
  {MacDonald}, \citenamefont {Tutuc},\ and\ \citenamefont
  {LeRoy}}]{PhysRevLett.121.037702}%
  \BibitemOpen
  \bibfield  {author} {\bibinfo {author} {\bibfnamefont {S.}~\bibnamefont
  {Huang}}, \bibinfo {author} {\bibfnamefont {K.}~\bibnamefont {Kim}}, \bibinfo
  {author} {\bibfnamefont {D.~K.}\ \bibnamefont {Efimkin}}, \bibinfo {author}
  {\bibfnamefont {T.}~\bibnamefont {Lovorn}}, \bibinfo {author} {\bibfnamefont
  {T.}~\bibnamefont {Taniguchi}}, \bibinfo {author} {\bibfnamefont
  {K.}~\bibnamefont {Watanabe}}, \bibinfo {author} {\bibfnamefont {A.~H.}\
  \bibnamefont {MacDonald}}, \bibinfo {author} {\bibfnamefont {E.}~\bibnamefont
  {Tutuc}}, \ and\ \bibinfo {author} {\bibfnamefont {B.~J.}\ \bibnamefont
  {LeRoy}},\ }\href {\doibase 10.1103/PhysRevLett.121.037702} {\bibfield
  {journal} {\bibinfo  {journal} {Phys. Rev. Lett.}\ }\textbf {\bibinfo
  {volume} {121}},\ \bibinfo {pages} {037702} (\bibinfo {year}
  {2018})}\BibitemShut {NoStop}%
\bibitem [{\citenamefont {Kerelsky}\ \emph {et~al.}(2019)\citenamefont
  {Kerelsky}, \citenamefont {McGilly}, \citenamefont {Kennes}, \citenamefont
  {Xian}, \citenamefont {Yankowitz}, \citenamefont {Chen}, \citenamefont
  {Watanabe}, \citenamefont {Taniguchi}, \citenamefont {Hone}, \citenamefont
  {Dean} \emph {et~al.}}]{kerelsky2019maximized}%
  \BibitemOpen
  \bibfield  {author} {\bibinfo {author} {\bibfnamefont {A.}~\bibnamefont
  {Kerelsky}}, \bibinfo {author} {\bibfnamefont {L.~J.}\ \bibnamefont
  {McGilly}}, \bibinfo {author} {\bibfnamefont {D.~M.}\ \bibnamefont {Kennes}},
  \bibinfo {author} {\bibfnamefont {L.}~\bibnamefont {Xian}}, \bibinfo {author}
  {\bibfnamefont {M.}~\bibnamefont {Yankowitz}}, \bibinfo {author}
  {\bibfnamefont {S.}~\bibnamefont {Chen}}, \bibinfo {author} {\bibfnamefont
  {K.}~\bibnamefont {Watanabe}}, \bibinfo {author} {\bibfnamefont
  {T.}~\bibnamefont {Taniguchi}}, \bibinfo {author} {\bibfnamefont
  {J.}~\bibnamefont {Hone}}, \bibinfo {author} {\bibfnamefont {C.}~\bibnamefont
  {Dean}},  \emph {et~al.},\ }\href@noop {} {\bibfield  {journal} {\bibinfo
  {journal} {Nature}\ }\textbf {\bibinfo {volume} {572}},\ \bibinfo {pages}
  {95} (\bibinfo {year} {2019})}\BibitemShut {NoStop}%
\bibitem [{\citenamefont {Xu}\ \emph {et~al.}(2019)\citenamefont {Xu},
  \citenamefont {Berdyugin}, \citenamefont {Kumaravadivel}, \citenamefont
  {Guinea}, \citenamefont {Krishna~Kumar}, \citenamefont {Bandurin},
  \citenamefont {Morozov}, \citenamefont {Kuang}, \citenamefont {Tsim},
  \citenamefont {Liu} \emph {et~al.}}]{xu2019giant}%
  \BibitemOpen
  \bibfield  {author} {\bibinfo {author} {\bibfnamefont {S.}~\bibnamefont
  {Xu}}, \bibinfo {author} {\bibfnamefont {A.}~\bibnamefont {Berdyugin}},
  \bibinfo {author} {\bibfnamefont {P.}~\bibnamefont {Kumaravadivel}}, \bibinfo
  {author} {\bibfnamefont {F.}~\bibnamefont {Guinea}}, \bibinfo {author}
  {\bibfnamefont {R.}~\bibnamefont {Krishna~Kumar}}, \bibinfo {author}
  {\bibfnamefont {D.}~\bibnamefont {Bandurin}}, \bibinfo {author}
  {\bibfnamefont {S.}~\bibnamefont {Morozov}}, \bibinfo {author} {\bibfnamefont
  {W.}~\bibnamefont {Kuang}}, \bibinfo {author} {\bibfnamefont
  {B.}~\bibnamefont {Tsim}}, \bibinfo {author} {\bibfnamefont {S.}~\bibnamefont
  {Liu}},  \emph {et~al.},\ }\href@noop {} {\bibfield  {journal} {\bibinfo
  {journal} {Nature communications}\ }\textbf {\bibinfo {volume} {10}},\
  \bibinfo {pages} {4008} (\bibinfo {year} {2019})}\BibitemShut {NoStop}%
\bibitem [{\citenamefont {Verbakel}\ \emph {et~al.}(2021)\citenamefont
  {Verbakel}, \citenamefont {Yao}, \citenamefont {Sotthewes},\ and\
  \citenamefont {Zandvliet}}]{verbakel2021valley}%
  \BibitemOpen
  \bibfield  {author} {\bibinfo {author} {\bibfnamefont {J.}~\bibnamefont
  {Verbakel}}, \bibinfo {author} {\bibfnamefont {Q.}~\bibnamefont {Yao}},
  \bibinfo {author} {\bibfnamefont {K.}~\bibnamefont {Sotthewes}}, \ and\
  \bibinfo {author} {\bibfnamefont {H.}~\bibnamefont {Zandvliet}},\ }\href@noop
  {} {\bibfield  {journal} {\bibinfo  {journal} {Physical Review B}\ }\textbf
  {\bibinfo {volume} {103}},\ \bibinfo {pages} {165134} (\bibinfo {year}
  {2021})}\BibitemShut {NoStop}%
\bibitem [{\citenamefont {Cao}\ \emph {et~al.}(2018{\natexlab{a}})\citenamefont
  {Cao}, \citenamefont {Fatemi}, \citenamefont {Fang}, \citenamefont
  {Watanabe}, \citenamefont {Taniguchi}, \citenamefont {Kaxiras},\ and\
  \citenamefont {Jarillo-Herrero}}]{cao2018unconventional}%
  \BibitemOpen
  \bibfield  {author} {\bibinfo {author} {\bibfnamefont {Y.}~\bibnamefont
  {Cao}}, \bibinfo {author} {\bibfnamefont {V.}~\bibnamefont {Fatemi}},
  \bibinfo {author} {\bibfnamefont {S.}~\bibnamefont {Fang}}, \bibinfo {author}
  {\bibfnamefont {K.}~\bibnamefont {Watanabe}}, \bibinfo {author}
  {\bibfnamefont {T.}~\bibnamefont {Taniguchi}}, \bibinfo {author}
  {\bibfnamefont {E.}~\bibnamefont {Kaxiras}}, \ and\ \bibinfo {author}
  {\bibfnamefont {P.}~\bibnamefont {Jarillo-Herrero}},\ }\href@noop {}
  {\bibfield  {journal} {\bibinfo  {journal} {Nature}\ }\textbf {\bibinfo
  {volume} {556}},\ \bibinfo {pages} {43} (\bibinfo {year}
  {2018}{\natexlab{a}})}\BibitemShut {NoStop}%
\bibitem [{\citenamefont {Cao}\ \emph {et~al.}(2018{\natexlab{b}})\citenamefont
  {Cao}, \citenamefont {Fatemi}, \citenamefont {Demir}, \citenamefont {Fang},
  \citenamefont {Tomarken}, \citenamefont {Luo}, \citenamefont
  {Sanchez-Yamagishi}, \citenamefont {Watanabe}, \citenamefont {Taniguchi},
  \citenamefont {Kaxiras} \emph {et~al.}}]{cao2018correlated}%
  \BibitemOpen
  \bibfield  {author} {\bibinfo {author} {\bibfnamefont {Y.}~\bibnamefont
  {Cao}}, \bibinfo {author} {\bibfnamefont {V.}~\bibnamefont {Fatemi}},
  \bibinfo {author} {\bibfnamefont {A.}~\bibnamefont {Demir}}, \bibinfo
  {author} {\bibfnamefont {S.}~\bibnamefont {Fang}}, \bibinfo {author}
  {\bibfnamefont {S.~L.}\ \bibnamefont {Tomarken}}, \bibinfo {author}
  {\bibfnamefont {J.~Y.}\ \bibnamefont {Luo}}, \bibinfo {author} {\bibfnamefont
  {J.~D.}\ \bibnamefont {Sanchez-Yamagishi}}, \bibinfo {author} {\bibfnamefont
  {K.}~\bibnamefont {Watanabe}}, \bibinfo {author} {\bibfnamefont
  {T.}~\bibnamefont {Taniguchi}}, \bibinfo {author} {\bibfnamefont
  {E.}~\bibnamefont {Kaxiras}},  \emph {et~al.},\ }\href@noop {} {\bibfield
  {journal} {\bibinfo  {journal} {Nature}\ }\textbf {\bibinfo {volume} {556}},\
  \bibinfo {pages} {80} (\bibinfo {year} {2018}{\natexlab{b}})}\BibitemShut
  {NoStop}%
\bibitem [{\citenamefont {Stepanov}\ \emph {et~al.}(2020)\citenamefont
  {Stepanov}, \citenamefont {Das}, \citenamefont {Lu}, \citenamefont
  {Fahimniya}, \citenamefont {Watanabe}, \citenamefont {Taniguchi},
  \citenamefont {Koppens}, \citenamefont {Lischner}, \citenamefont {Levitov},\
  and\ \citenamefont {Efetov}}]{stepanov2020untying}%
  \BibitemOpen
  \bibfield  {author} {\bibinfo {author} {\bibfnamefont {P.}~\bibnamefont
  {Stepanov}}, \bibinfo {author} {\bibfnamefont {I.}~\bibnamefont {Das}},
  \bibinfo {author} {\bibfnamefont {X.}~\bibnamefont {Lu}}, \bibinfo {author}
  {\bibfnamefont {A.}~\bibnamefont {Fahimniya}}, \bibinfo {author}
  {\bibfnamefont {K.}~\bibnamefont {Watanabe}}, \bibinfo {author}
  {\bibfnamefont {T.}~\bibnamefont {Taniguchi}}, \bibinfo {author}
  {\bibfnamefont {F.~H.}\ \bibnamefont {Koppens}}, \bibinfo {author}
  {\bibfnamefont {J.}~\bibnamefont {Lischner}}, \bibinfo {author}
  {\bibfnamefont {L.}~\bibnamefont {Levitov}}, \ and\ \bibinfo {author}
  {\bibfnamefont {D.~K.}\ \bibnamefont {Efetov}},\ }\href@noop {} {\bibfield
  {journal} {\bibinfo  {journal} {Nature}\ }\textbf {\bibinfo {volume} {583}},\
  \bibinfo {pages} {375} (\bibinfo {year} {2020})}\BibitemShut {NoStop}%
\bibitem [{\citenamefont {Popov}\ \emph {et~al.}(2011)\citenamefont {Popov},
  \citenamefont {Lebedeva}, \citenamefont {Knizhnik}, \citenamefont {Lozovik},\
  and\ \citenamefont {Potapkin}}]{popov2011commensurate}%
  \BibitemOpen
  \bibfield  {author} {\bibinfo {author} {\bibfnamefont {A.~M.}\ \bibnamefont
  {Popov}}, \bibinfo {author} {\bibfnamefont {I.~V.}\ \bibnamefont {Lebedeva}},
  \bibinfo {author} {\bibfnamefont {A.~A.}\ \bibnamefont {Knizhnik}}, \bibinfo
  {author} {\bibfnamefont {Y.~E.}\ \bibnamefont {Lozovik}}, \ and\ \bibinfo
  {author} {\bibfnamefont {B.~V.}\ \bibnamefont {Potapkin}},\ }\href@noop {}
  {\bibfield  {journal} {\bibinfo  {journal} {Physical Review B}\ }\textbf
  {\bibinfo {volume} {84}},\ \bibinfo {pages} {045404} (\bibinfo {year}
  {2011})}\BibitemShut {NoStop}%
\bibitem [{\citenamefont {Alden}\ \emph {et~al.}(2013)\citenamefont {Alden},
  \citenamefont {Tsen}, \citenamefont {Huang}, \citenamefont {Hovden},
  \citenamefont {Brown}, \citenamefont {Park}, \citenamefont {Muller},\ and\
  \citenamefont {McEuen}}]{alden2013strain}%
  \BibitemOpen
  \bibfield  {author} {\bibinfo {author} {\bibfnamefont {J.~S.}\ \bibnamefont
  {Alden}}, \bibinfo {author} {\bibfnamefont {A.~W.}\ \bibnamefont {Tsen}},
  \bibinfo {author} {\bibfnamefont {P.~Y.}\ \bibnamefont {Huang}}, \bibinfo
  {author} {\bibfnamefont {R.}~\bibnamefont {Hovden}}, \bibinfo {author}
  {\bibfnamefont {L.}~\bibnamefont {Brown}}, \bibinfo {author} {\bibfnamefont
  {J.}~\bibnamefont {Park}}, \bibinfo {author} {\bibfnamefont {D.~A.}\
  \bibnamefont {Muller}}, \ and\ \bibinfo {author} {\bibfnamefont {P.~L.}\
  \bibnamefont {McEuen}},\ }\href@noop {} {\bibfield  {journal} {\bibinfo
  {journal} {Proceedings of the National Academy of Sciences}\ }\textbf
  {\bibinfo {volume} {110}},\ \bibinfo {pages} {11256} (\bibinfo {year}
  {2013})}\BibitemShut {NoStop}%
\bibitem [{\citenamefont {Choi}\ \emph {et~al.}(2010)\citenamefont {Choi},
  \citenamefont {Jhi},\ and\ \citenamefont {Son}}]{choi2010controlling}%
  \BibitemOpen
  \bibfield  {author} {\bibinfo {author} {\bibfnamefont {S.-M.}\ \bibnamefont
  {Choi}}, \bibinfo {author} {\bibfnamefont {S.-H.}\ \bibnamefont {Jhi}}, \
  and\ \bibinfo {author} {\bibfnamefont {Y.-W.}\ \bibnamefont {Son}},\
  }\href@noop {} {\bibfield  {journal} {\bibinfo  {journal} {Nano letters}\
  }\textbf {\bibinfo {volume} {10}},\ \bibinfo {pages} {3486} (\bibinfo {year}
  {2010})}\BibitemShut {NoStop}%
\bibitem [{\citenamefont {Van~der Donck}\ \emph {et~al.}(2016)\citenamefont
  {Van~der Donck}, \citenamefont {De~Beule}, \citenamefont {Partoens},
  \citenamefont {Peeters},\ and\ \citenamefont
  {Van~Duppen}}]{van2016piezoelectricity}%
  \BibitemOpen
  \bibfield  {author} {\bibinfo {author} {\bibfnamefont {M.}~\bibnamefont
  {Van~der Donck}}, \bibinfo {author} {\bibfnamefont {C.}~\bibnamefont
  {De~Beule}}, \bibinfo {author} {\bibfnamefont {B.}~\bibnamefont {Partoens}},
  \bibinfo {author} {\bibfnamefont {F.}~\bibnamefont {Peeters}}, \ and\
  \bibinfo {author} {\bibfnamefont {B.}~\bibnamefont {Van~Duppen}},\
  }\href@noop {} {\bibfield  {journal} {\bibinfo  {journal} {2D Materials}\
  }\textbf {\bibinfo {volume} {3}},\ \bibinfo {pages} {035015} (\bibinfo {year}
  {2016})}\BibitemShut {NoStop}%
\bibitem [{\citenamefont {Georgoulea}\ \emph {et~al.}(2022)\citenamefont
  {Georgoulea}, \citenamefont {Power},\ and\ \citenamefont
  {Caffrey}}]{georgoulea2022strain}%
  \BibitemOpen
  \bibfield  {author} {\bibinfo {author} {\bibfnamefont {N.~C.}\ \bibnamefont
  {Georgoulea}}, \bibinfo {author} {\bibfnamefont {S.~R.}\ \bibnamefont
  {Power}}, \ and\ \bibinfo {author} {\bibfnamefont {N.~M.}\ \bibnamefont
  {Caffrey}},\ }\href@noop {} {\bibfield  {journal} {\bibinfo  {journal}
  {Journal of Physics: Condensed Matter}\ }\textbf {\bibinfo {volume} {34}},\
  \bibinfo {pages} {475302} (\bibinfo {year} {2022})}\BibitemShut {NoStop}%
\bibitem [{\citenamefont {Schleder}\ \emph {et~al.}(2023)\citenamefont
  {Schleder}, \citenamefont {Pizzochero},\ and\ \citenamefont
  {Kaxiras}}]{schleder2023onedimensional}%
  \BibitemOpen
  \bibfield  {author} {\bibinfo {author} {\bibfnamefont {G.~R.}\ \bibnamefont
  {Schleder}}, \bibinfo {author} {\bibfnamefont {M.}~\bibnamefont
  {Pizzochero}}, \ and\ \bibinfo {author} {\bibfnamefont {E.}~\bibnamefont
  {Kaxiras}},\ }\href@noop {} {\enquote {\bibinfo {title} {One-dimensional
  moir\'e physics and chemistry in heterostrained bilayer graphene},}\ }
  (\bibinfo {year} {2023}),\ \Eprint {http://arxiv.org/abs/2306.09799}
  {arXiv:2306.09799 [cond-mat.mtrl-sci]} \BibitemShut {NoStop}%
\bibitem [{\citenamefont {Androulidakis}\ \emph {et~al.}(2020)\citenamefont
  {Androulidakis}, \citenamefont {Koukaras}, \citenamefont {Paterakis},
  \citenamefont {Trakakis},\ and\ \citenamefont
  {Galiotis}}]{androulidakis2020tunable}%
  \BibitemOpen
  \bibfield  {author} {\bibinfo {author} {\bibfnamefont {C.}~\bibnamefont
  {Androulidakis}}, \bibinfo {author} {\bibfnamefont {E.~N.}\ \bibnamefont
  {Koukaras}}, \bibinfo {author} {\bibfnamefont {G.}~\bibnamefont {Paterakis}},
  \bibinfo {author} {\bibfnamefont {G.}~\bibnamefont {Trakakis}}, \ and\
  \bibinfo {author} {\bibfnamefont {C.}~\bibnamefont {Galiotis}},\ }\href@noop
  {} {\bibfield  {journal} {\bibinfo  {journal} {Nature communications}\
  }\textbf {\bibinfo {volume} {11}},\ \bibinfo {pages} {1595} (\bibinfo {year}
  {2020})}\BibitemShut {NoStop}%
\bibitem [{\citenamefont {Frank}\ \emph {et~al.}(2012)\citenamefont {Frank},
  \citenamefont {Bousa}, \citenamefont {Riaz}, \citenamefont {Jalil},
  \citenamefont {Novoselov}, \citenamefont {Tsoukleri}, \citenamefont
  {Parthenios}, \citenamefont {Kavan}, \citenamefont {Papagelis},\ and\
  \citenamefont {Galiotis}}]{frank2012phonon}%
  \BibitemOpen
  \bibfield  {author} {\bibinfo {author} {\bibfnamefont {O.}~\bibnamefont
  {Frank}}, \bibinfo {author} {\bibfnamefont {M.}~\bibnamefont {Bousa}},
  \bibinfo {author} {\bibfnamefont {I.}~\bibnamefont {Riaz}}, \bibinfo {author}
  {\bibfnamefont {R.}~\bibnamefont {Jalil}}, \bibinfo {author} {\bibfnamefont
  {K.~S.}\ \bibnamefont {Novoselov}}, \bibinfo {author} {\bibfnamefont
  {G.}~\bibnamefont {Tsoukleri}}, \bibinfo {author} {\bibfnamefont
  {J.}~\bibnamefont {Parthenios}}, \bibinfo {author} {\bibfnamefont
  {L.}~\bibnamefont {Kavan}}, \bibinfo {author} {\bibfnamefont
  {K.}~\bibnamefont {Papagelis}}, \ and\ \bibinfo {author} {\bibfnamefont
  {C.}~\bibnamefont {Galiotis}},\ }\href@noop {} {\bibfield  {journal}
  {\bibinfo  {journal} {Nano Letters}\ }\textbf {\bibinfo {volume} {12}},\
  \bibinfo {pages} {687} (\bibinfo {year} {2012})}\BibitemShut {NoStop}%
\bibitem [{\citenamefont {Wang}\ \emph {et~al.}(2019)\citenamefont {Wang},
  \citenamefont {Ouyang}, \citenamefont {Cao}, \citenamefont {Ma},\ and\
  \citenamefont {Zheng}}]{wang2019robust}%
  \BibitemOpen
  \bibfield  {author} {\bibinfo {author} {\bibfnamefont {K.}~\bibnamefont
  {Wang}}, \bibinfo {author} {\bibfnamefont {W.}~\bibnamefont {Ouyang}},
  \bibinfo {author} {\bibfnamefont {W.}~\bibnamefont {Cao}}, \bibinfo {author}
  {\bibfnamefont {M.}~\bibnamefont {Ma}}, \ and\ \bibinfo {author}
  {\bibfnamefont {Q.}~\bibnamefont {Zheng}},\ }\href@noop {} {\bibfield
  {journal} {\bibinfo  {journal} {Nanoscale}\ }\textbf {\bibinfo {volume}
  {11}},\ \bibinfo {pages} {2186} (\bibinfo {year} {2019})}\BibitemShut
  {NoStop}%
\bibitem [{\citenamefont {Lin}\ \emph {et~al.}(2013)\citenamefont {Lin},
  \citenamefont {Fang}, \citenamefont {Zhou}, \citenamefont {Lupini},
  \citenamefont {Idrobo}, \citenamefont {Kong}, \citenamefont {Pennycook},\
  and\ \citenamefont {Pantelides}}]{lin2013ac}%
  \BibitemOpen
  \bibfield  {author} {\bibinfo {author} {\bibfnamefont {J.}~\bibnamefont
  {Lin}}, \bibinfo {author} {\bibfnamefont {W.}~\bibnamefont {Fang}}, \bibinfo
  {author} {\bibfnamefont {W.}~\bibnamefont {Zhou}}, \bibinfo {author}
  {\bibfnamefont {A.~R.}\ \bibnamefont {Lupini}}, \bibinfo {author}
  {\bibfnamefont {J.~C.}\ \bibnamefont {Idrobo}}, \bibinfo {author}
  {\bibfnamefont {J.}~\bibnamefont {Kong}}, \bibinfo {author} {\bibfnamefont
  {S.~J.}\ \bibnamefont {Pennycook}}, \ and\ \bibinfo {author} {\bibfnamefont
  {S.~T.}\ \bibnamefont {Pantelides}},\ }\href@noop {} {\bibfield  {journal}
  {\bibinfo  {journal} {Nano letters}\ }\textbf {\bibinfo {volume} {13}},\
  \bibinfo {pages} {3262} (\bibinfo {year} {2013})}\BibitemShut {NoStop}%
\bibitem [{\citenamefont {Van~Wijk}\ \emph {et~al.}(2015)\citenamefont
  {Van~Wijk}, \citenamefont {Schuring}, \citenamefont {Katsnelson},\ and\
  \citenamefont {Fasolino}}]{van2015relaxation}%
  \BibitemOpen
  \bibfield  {author} {\bibinfo {author} {\bibfnamefont {M.}~\bibnamefont
  {Van~Wijk}}, \bibinfo {author} {\bibfnamefont {A.}~\bibnamefont {Schuring}},
  \bibinfo {author} {\bibfnamefont {M.}~\bibnamefont {Katsnelson}}, \ and\
  \bibinfo {author} {\bibfnamefont {A.}~\bibnamefont {Fasolino}},\ }\href@noop
  {} {\bibfield  {journal} {\bibinfo  {journal} {2D Materials}\ }\textbf
  {\bibinfo {volume} {2}},\ \bibinfo {pages} {034010} (\bibinfo {year}
  {2015})}\BibitemShut {NoStop}%
\bibitem [{\citenamefont {Dai}\ \emph {et~al.}(2016)\citenamefont {Dai},
  \citenamefont {Xiang},\ and\ \citenamefont {Srolovitz}}]{dai2016twisted}%
  \BibitemOpen
  \bibfield  {author} {\bibinfo {author} {\bibfnamefont {S.}~\bibnamefont
  {Dai}}, \bibinfo {author} {\bibfnamefont {Y.}~\bibnamefont {Xiang}}, \ and\
  \bibinfo {author} {\bibfnamefont {D.~J.}\ \bibnamefont {Srolovitz}},\
  }\href@noop {} {\bibfield  {journal} {\bibinfo  {journal} {Nano letters}\
  }\textbf {\bibinfo {volume} {16}},\ \bibinfo {pages} {5923} (\bibinfo {year}
  {2016})}\BibitemShut {NoStop}%
\bibitem [{\citenamefont {Jain}\ \emph {et~al.}(2016)\citenamefont {Jain},
  \citenamefont {Juri{\v{c}}i{\'c}},\ and\ \citenamefont
  {Barkema}}]{jain2016structure}%
  \BibitemOpen
  \bibfield  {author} {\bibinfo {author} {\bibfnamefont {S.~K.}\ \bibnamefont
  {Jain}}, \bibinfo {author} {\bibfnamefont {V.}~\bibnamefont
  {Juri{\v{c}}i{\'c}}}, \ and\ \bibinfo {author} {\bibfnamefont {G.~T.}\
  \bibnamefont {Barkema}},\ }\href@noop {} {\bibfield  {journal} {\bibinfo
  {journal} {2D Materials}\ }\textbf {\bibinfo {volume} {4}},\ \bibinfo {pages}
  {015018} (\bibinfo {year} {2016})}\BibitemShut {NoStop}%
\bibitem [{\citenamefont {Nam}\ and\ \citenamefont
  {Koshino}(2017)}]{nam2017lattice}%
  \BibitemOpen
  \bibfield  {author} {\bibinfo {author} {\bibfnamefont {N.~N.}\ \bibnamefont
  {Nam}}\ and\ \bibinfo {author} {\bibfnamefont {M.}~\bibnamefont {Koshino}},\
  }\href@noop {} {\bibfield  {journal} {\bibinfo  {journal} {Physical Review
  B}\ }\textbf {\bibinfo {volume} {96}},\ \bibinfo {pages} {075311} (\bibinfo
  {year} {2017})}\BibitemShut {NoStop}%
\bibitem [{\citenamefont {Gargiulo}\ and\ \citenamefont
  {Yazyev}(2017)}]{gargiulo2017structural}%
  \BibitemOpen
  \bibfield  {author} {\bibinfo {author} {\bibfnamefont {F.}~\bibnamefont
  {Gargiulo}}\ and\ \bibinfo {author} {\bibfnamefont {O.~V.}\ \bibnamefont
  {Yazyev}},\ }\href@noop {} {\bibfield  {journal} {\bibinfo  {journal} {2D
  Materials}\ }\textbf {\bibinfo {volume} {5}},\ \bibinfo {pages} {015019}
  (\bibinfo {year} {2017})}\BibitemShut {NoStop}%
\bibitem [{\citenamefont {Carr}\ \emph {et~al.}(2018)\citenamefont {Carr},
  \citenamefont {Massatt}, \citenamefont {Torrisi}, \citenamefont {Cazeaux},
  \citenamefont {Luskin},\ and\ \citenamefont {Kaxiras}}]{carr2018relaxation}%
  \BibitemOpen
  \bibfield  {author} {\bibinfo {author} {\bibfnamefont {S.}~\bibnamefont
  {Carr}}, \bibinfo {author} {\bibfnamefont {D.}~\bibnamefont {Massatt}},
  \bibinfo {author} {\bibfnamefont {S.~B.}\ \bibnamefont {Torrisi}}, \bibinfo
  {author} {\bibfnamefont {P.}~\bibnamefont {Cazeaux}}, \bibinfo {author}
  {\bibfnamefont {M.}~\bibnamefont {Luskin}}, \ and\ \bibinfo {author}
  {\bibfnamefont {E.}~\bibnamefont {Kaxiras}},\ }\href@noop {} {\bibfield
  {journal} {\bibinfo  {journal} {Physical Review B}\ }\textbf {\bibinfo
  {volume} {98}},\ \bibinfo {pages} {224102} (\bibinfo {year}
  {2018})}\BibitemShut {NoStop}%
\bibitem [{\citenamefont {Lin}\ \emph {et~al.}(2018)\citenamefont {Lin},
  \citenamefont {Liu},\ and\ \citenamefont {Tom{\'a}nek}}]{lin2018shear}%
  \BibitemOpen
  \bibfield  {author} {\bibinfo {author} {\bibfnamefont {X.}~\bibnamefont
  {Lin}}, \bibinfo {author} {\bibfnamefont {D.}~\bibnamefont {Liu}}, \ and\
  \bibinfo {author} {\bibfnamefont {D.}~\bibnamefont {Tom{\'a}nek}},\
  }\href@noop {} {\bibfield  {journal} {\bibinfo  {journal} {Physical Review
  B}\ }\textbf {\bibinfo {volume} {98}},\ \bibinfo {pages} {195432} (\bibinfo
  {year} {2018})}\BibitemShut {NoStop}%
\bibitem [{\citenamefont {Guinea}\ and\ \citenamefont
  {Walet}(2019)}]{guinea2019continuum}%
  \BibitemOpen
  \bibfield  {author} {\bibinfo {author} {\bibfnamefont {F.}~\bibnamefont
  {Guinea}}\ and\ \bibinfo {author} {\bibfnamefont {N.~R.}\ \bibnamefont
  {Walet}},\ }\href@noop {} {\bibfield  {journal} {\bibinfo  {journal}
  {Physical Review B}\ }\textbf {\bibinfo {volume} {99}},\ \bibinfo {pages}
  {205134} (\bibinfo {year} {2019})}\BibitemShut {NoStop}%
\bibitem [{\citenamefont {Yoo}\ \emph {et~al.}(2019)\citenamefont {Yoo},
  \citenamefont {Engelke}, \citenamefont {Carr}, \citenamefont {Fang},
  \citenamefont {Zhang}, \citenamefont {Cazeaux}, \citenamefont {Sung},
  \citenamefont {Hovden}, \citenamefont {Tsen}, \citenamefont {Taniguchi} \emph
  {et~al.}}]{yoo2019atomic}%
  \BibitemOpen
  \bibfield  {author} {\bibinfo {author} {\bibfnamefont {H.}~\bibnamefont
  {Yoo}}, \bibinfo {author} {\bibfnamefont {R.}~\bibnamefont {Engelke}},
  \bibinfo {author} {\bibfnamefont {S.}~\bibnamefont {Carr}}, \bibinfo {author}
  {\bibfnamefont {S.}~\bibnamefont {Fang}}, \bibinfo {author} {\bibfnamefont
  {K.}~\bibnamefont {Zhang}}, \bibinfo {author} {\bibfnamefont
  {P.}~\bibnamefont {Cazeaux}}, \bibinfo {author} {\bibfnamefont {S.~H.}\
  \bibnamefont {Sung}}, \bibinfo {author} {\bibfnamefont {R.}~\bibnamefont
  {Hovden}}, \bibinfo {author} {\bibfnamefont {A.~W.}\ \bibnamefont {Tsen}},
  \bibinfo {author} {\bibfnamefont {T.}~\bibnamefont {Taniguchi}},  \emph
  {et~al.},\ }\href@noop {} {\bibfield  {journal} {\bibinfo  {journal} {Nature
  materials}\ }\textbf {\bibinfo {volume} {18}},\ \bibinfo {pages} {448}
  (\bibinfo {year} {2019})}\BibitemShut {NoStop}%
\bibitem [{\citenamefont {Lucignano}\ \emph {et~al.}(2019)\citenamefont
  {Lucignano}, \citenamefont {Alf{\`e}}, \citenamefont {Cataudella},
  \citenamefont {Ninno},\ and\ \citenamefont {Cantele}}]{lucignano2019crucial}%
  \BibitemOpen
  \bibfield  {author} {\bibinfo {author} {\bibfnamefont {P.}~\bibnamefont
  {Lucignano}}, \bibinfo {author} {\bibfnamefont {D.}~\bibnamefont {Alf{\`e}}},
  \bibinfo {author} {\bibfnamefont {V.}~\bibnamefont {Cataudella}}, \bibinfo
  {author} {\bibfnamefont {D.}~\bibnamefont {Ninno}}, \ and\ \bibinfo {author}
  {\bibfnamefont {G.}~\bibnamefont {Cantele}},\ }\href@noop {} {\bibfield
  {journal} {\bibinfo  {journal} {Physical Review B}\ }\textbf {\bibinfo
  {volume} {99}},\ \bibinfo {pages} {195419} (\bibinfo {year}
  {2019})}\BibitemShut {NoStop}%
\bibitem [{\citenamefont {Nguyen}\ \emph {et~al.}(2021)\citenamefont {Nguyen},
  \citenamefont {Paszko}, \citenamefont {Lamparski}, \citenamefont
  {Van~Troeye}, \citenamefont {Meunier},\ and\ \citenamefont
  {Charlier}}]{nguyen2021electronic}%
  \BibitemOpen
  \bibfield  {author} {\bibinfo {author} {\bibfnamefont {V.~H.}\ \bibnamefont
  {Nguyen}}, \bibinfo {author} {\bibfnamefont {D.}~\bibnamefont {Paszko}},
  \bibinfo {author} {\bibfnamefont {M.}~\bibnamefont {Lamparski}}, \bibinfo
  {author} {\bibfnamefont {B.}~\bibnamefont {Van~Troeye}}, \bibinfo {author}
  {\bibfnamefont {V.}~\bibnamefont {Meunier}}, \ and\ \bibinfo {author}
  {\bibfnamefont {J.-C.}\ \bibnamefont {Charlier}},\ }\href@noop {} {\bibfield
  {journal} {\bibinfo  {journal} {2D Materials}\ }\textbf {\bibinfo {volume}
  {8}},\ \bibinfo {pages} {035046} (\bibinfo {year} {2021})}\BibitemShut
  {NoStop}%
\bibitem [{\citenamefont {Leconte}\ \emph {et~al.}(2022)\citenamefont
  {Leconte}, \citenamefont {Javvaji}, \citenamefont {An}, \citenamefont
  {Samudrala},\ and\ \citenamefont {Jung}}]{leconte2022relaxation}%
  \BibitemOpen
  \bibfield  {author} {\bibinfo {author} {\bibfnamefont {N.}~\bibnamefont
  {Leconte}}, \bibinfo {author} {\bibfnamefont {S.}~\bibnamefont {Javvaji}},
  \bibinfo {author} {\bibfnamefont {J.}~\bibnamefont {An}}, \bibinfo {author}
  {\bibfnamefont {A.}~\bibnamefont {Samudrala}}, \ and\ \bibinfo {author}
  {\bibfnamefont {J.}~\bibnamefont {Jung}},\ }\href {\doibase
  10.1103/PhysRevB.106.115410} {\bibfield  {journal} {\bibinfo  {journal}
  {Phys. Rev. B}\ }\textbf {\bibinfo {volume} {106}},\ \bibinfo {pages}
  {115410} (\bibinfo {year} {2022})}\BibitemShut {NoStop}%
\bibitem [{\citenamefont {Yu}\ \emph {et~al.}(2008)\citenamefont {Yu},
  \citenamefont {Ni}, \citenamefont {Du}, \citenamefont {You}, \citenamefont
  {Wang},\ and\ \citenamefont {Shen}}]{yu2008raman}%
  \BibitemOpen
  \bibfield  {author} {\bibinfo {author} {\bibfnamefont {T.}~\bibnamefont
  {Yu}}, \bibinfo {author} {\bibfnamefont {Z.}~\bibnamefont {Ni}}, \bibinfo
  {author} {\bibfnamefont {C.}~\bibnamefont {Du}}, \bibinfo {author}
  {\bibfnamefont {Y.}~\bibnamefont {You}}, \bibinfo {author} {\bibfnamefont
  {Y.}~\bibnamefont {Wang}}, \ and\ \bibinfo {author} {\bibfnamefont
  {Z.}~\bibnamefont {Shen}},\ }\href@noop {} {\bibfield  {journal} {\bibinfo
  {journal} {The Journal of Physical Chemistry C}\ }\textbf {\bibinfo {volume}
  {112}},\ \bibinfo {pages} {12602} (\bibinfo {year} {2008})}\BibitemShut
  {NoStop}%
\bibitem [{\citenamefont {Huang}\ \emph {et~al.}(2009)\citenamefont {Huang},
  \citenamefont {Yan}, \citenamefont {Chen}, \citenamefont {Song},
  \citenamefont {Heinz},\ and\ \citenamefont {Hone}}]{huang2009phonon}%
  \BibitemOpen
  \bibfield  {author} {\bibinfo {author} {\bibfnamefont {M.}~\bibnamefont
  {Huang}}, \bibinfo {author} {\bibfnamefont {H.}~\bibnamefont {Yan}}, \bibinfo
  {author} {\bibfnamefont {C.}~\bibnamefont {Chen}}, \bibinfo {author}
  {\bibfnamefont {D.}~\bibnamefont {Song}}, \bibinfo {author} {\bibfnamefont
  {T.~F.}\ \bibnamefont {Heinz}}, \ and\ \bibinfo {author} {\bibfnamefont
  {J.}~\bibnamefont {Hone}},\ }\href@noop {} {\bibfield  {journal} {\bibinfo
  {journal} {Proceedings of the National Academy of Sciences}\ }\textbf
  {\bibinfo {volume} {106}},\ \bibinfo {pages} {7304} (\bibinfo {year}
  {2009})}\BibitemShut {NoStop}%
\bibitem [{\citenamefont {Anagnostopoulos}\ \emph {et~al.}(2015)\citenamefont
  {Anagnostopoulos}, \citenamefont {Androulidakis}, \citenamefont {Koukaras},
  \citenamefont {Tsoukleri}, \citenamefont {Polyzos}, \citenamefont
  {Parthenios}, \citenamefont {Papagelis},\ and\ \citenamefont
  {Galiotis}}]{anagnostopoulos2015stress}%
  \BibitemOpen
  \bibfield  {author} {\bibinfo {author} {\bibfnamefont {G.}~\bibnamefont
  {Anagnostopoulos}}, \bibinfo {author} {\bibfnamefont {C.}~\bibnamefont
  {Androulidakis}}, \bibinfo {author} {\bibfnamefont {E.~N.}\ \bibnamefont
  {Koukaras}}, \bibinfo {author} {\bibfnamefont {G.}~\bibnamefont {Tsoukleri}},
  \bibinfo {author} {\bibfnamefont {I.}~\bibnamefont {Polyzos}}, \bibinfo
  {author} {\bibfnamefont {J.}~\bibnamefont {Parthenios}}, \bibinfo {author}
  {\bibfnamefont {K.}~\bibnamefont {Papagelis}}, \ and\ \bibinfo {author}
  {\bibfnamefont {C.}~\bibnamefont {Galiotis}},\ }\href@noop {} {\bibfield
  {journal} {\bibinfo  {journal} {ACS applied materials \& interfaces}\
  }\textbf {\bibinfo {volume} {7}},\ \bibinfo {pages} {4216} (\bibinfo {year}
  {2015})}\BibitemShut {NoStop}%
\bibitem [{\citenamefont {Kapfer}\ \emph {et~al.}(2022)\citenamefont {Kapfer},
  \citenamefont {Jessen}, \citenamefont {Eisele}, \citenamefont {Fu},
  \citenamefont {Danielsen}, \citenamefont {Darlington}, \citenamefont {Moore},
  \citenamefont {Finney}, \citenamefont {Marchese}, \citenamefont {Hsieh} \emph
  {et~al.}}]{kapfer2022programming}%
  \BibitemOpen
  \bibfield  {author} {\bibinfo {author} {\bibfnamefont {M.}~\bibnamefont
  {Kapfer}}, \bibinfo {author} {\bibfnamefont {B.~S.}\ \bibnamefont {Jessen}},
  \bibinfo {author} {\bibfnamefont {M.~E.}\ \bibnamefont {Eisele}}, \bibinfo
  {author} {\bibfnamefont {M.}~\bibnamefont {Fu}}, \bibinfo {author}
  {\bibfnamefont {D.~R.}\ \bibnamefont {Danielsen}}, \bibinfo {author}
  {\bibfnamefont {T.~P.}\ \bibnamefont {Darlington}}, \bibinfo {author}
  {\bibfnamefont {S.~L.}\ \bibnamefont {Moore}}, \bibinfo {author}
  {\bibfnamefont {N.~R.}\ \bibnamefont {Finney}}, \bibinfo {author}
  {\bibfnamefont {A.}~\bibnamefont {Marchese}}, \bibinfo {author}
  {\bibfnamefont {V.}~\bibnamefont {Hsieh}},  \emph {et~al.},\ }\href@noop {}
  {\bibfield  {journal} {\bibinfo  {journal} {arXiv preprint arXiv:2209.10696}\
  } (\bibinfo {year} {2022})}\BibitemShut {NoStop}%
\bibitem [{\citenamefont {Slater}\ and\ \citenamefont
  {Koster}(1954)}]{slater1954simplified}%
  \BibitemOpen
  \bibfield  {author} {\bibinfo {author} {\bibfnamefont {J.~C.}\ \bibnamefont
  {Slater}}\ and\ \bibinfo {author} {\bibfnamefont {G.~F.}\ \bibnamefont
  {Koster}},\ }\href@noop {} {\bibfield  {journal} {\bibinfo  {journal}
  {Physical Review}\ }\textbf {\bibinfo {volume} {94}},\ \bibinfo {pages}
  {1498} (\bibinfo {year} {1954})}\BibitemShut {NoStop}%
\bibitem [{\citenamefont {Trambly~de Laissardi{\`e}re}\ \emph
  {et~al.}(2010)\citenamefont {Trambly~de Laissardi{\`e}re}, \citenamefont
  {Mayou},\ and\ \citenamefont {Magaud}}]{trambly2010localization}%
  \BibitemOpen
  \bibfield  {author} {\bibinfo {author} {\bibfnamefont {G.}~\bibnamefont
  {Trambly~de Laissardi{\`e}re}}, \bibinfo {author} {\bibfnamefont
  {D.}~\bibnamefont {Mayou}}, \ and\ \bibinfo {author} {\bibfnamefont
  {L.}~\bibnamefont {Magaud}},\ }\href@noop {} {\bibfield  {journal} {\bibinfo
  {journal} {Nano letters}\ }\textbf {\bibinfo {volume} {10}},\ \bibinfo
  {pages} {804} (\bibinfo {year} {2010})}\BibitemShut {NoStop}%
\bibitem [{\citenamefont {Parhizkar}\ and\ \citenamefont
  {Galitski}(2022)}]{parhizkar2022strained}%
  \BibitemOpen
  \bibfield  {author} {\bibinfo {author} {\bibfnamefont {A.}~\bibnamefont
  {Parhizkar}}\ and\ \bibinfo {author} {\bibfnamefont {V.}~\bibnamefont
  {Galitski}},\ }\href {\doibase 10.1103/PhysRevResearch.4.L022027} {\bibfield
  {journal} {\bibinfo  {journal} {Phys. Rev. Res.}\ }\textbf {\bibinfo {volume}
  {4}},\ \bibinfo {pages} {L022027} (\bibinfo {year} {2022})}\BibitemShut
  {NoStop}%
\end{thebibliography}%
\end{document}